\documentclass[aps,twocolumn,preprintnumbers,nofootinbib,superscriptaddress]{revtex4}
\usepackage{amsmath}
\usepackage{graphicx}
\usepackage{amsfonts}
\usepackage{array}
\usepackage{amsthm}
\usepackage{bm}
\usepackage{supertabular}
\usepackage{subfig}
\usepackage[breaklinks]{hyperref}
\usepackage{color}
\usepackage[font=small,labelfont=bf,
justification=justified,
format=plain]{caption}
\usepackage{graphicx}
\usepackage{caption}
\usepackage[font=small,labelfont=bf,justification=justified]{caption}
\usepackage[labelfont=bf,labelsep=quad,justification=justified]{caption}
\usepackage{epstopdf}

\newcommand{\be}{\begin{equation}}
\newcommand{\ee}{\end{equation}}
\newcommand{\ba}{\begin{eqnarray}}
\newcommand{\ea}{\end{eqnarray}}
\newcommand{\bal}{\begin{align}}
\newcommand{\eal}{\end{align}}

\newcommand{\bw}{\begin{widetext}}
\newcommand{\ew}{\end{widetext}}

\def\d{\partial}

\def\a{\alpha'}

\def\beq{\begin{equation}}

\def\eeq{\end{equation}}

\relax

\begin{document}

\title{Theory and phenomenology of a four dimensional string--corrected black hole}
\author{Kimet Jusufi}
\email{kimet.jusufi@unite.edu.mk}
\affiliation{Physics Department, State University of Tetovo, Ilinden Street nn, 1200,
Tetovo, North Macedonia}
\author{Dejan Stojkovic}
\email{ds77@buffalo.edu}
\affiliation{HEPCOS, Department of Physics, SUNY at Buffalo, Buffalo, NY 14260-1500, U.S.A.}

\begin{abstract}
We construct  an effective four dimensional string-corrected black hole (4D SCBH) by rescaling the string coupling parameter in a $D$-dimensional Callan-Myers-Perry black hole. From the theoretical point of view, the most interesting findings are that the string corrections coincide with the so-called generalized uncertainty principle (GUP) corrections to black hole solutions, Bekenstein-Hawking entropy acquires logarithmic corrections, and that there exists a critical value of the coupling parameter for which the black hole temperature vanishes. We also find that, due to the string corrections the nature of the central singularity may be altered from spacelike to timelike singularity. In addition, we study the possibility of testing such a black hole with astrophysical observations. Since the dilaton field does not decouple from the metric it is not a priori clear that the resulting 4D SCBH offers only small corrections to the Schwarzschild black hole.
We used motion of the S2 star around the black hole at the center of our galaxy to constrain the parameters (the string coupling parameter and ADM mass) of the 4D SCBH. To test the weak gravity regime we calculate the deflection angle in this geometry and apply it to gravitational lensing. To test the strong field regime, we calculate the black hole shadow radius. While we find that the observables change as we change the string coupling parameter, the magnitude of the change is too small to distinguish it from the Schwarzschild black hole. With the current precision, to the leading order terms, the 4D SCBH cannot be distinguished from the Schwarzschild black hole.   
\end{abstract}

\maketitle

\section{Introduction}
Black holes (BHs), as the most interesting solution of Einstein's equations, are unique objects that can provide us with both theoretical and phenomenological tests of our new models. Today we have numerous ways to test black hole physics on astrophysical scales in both weak and strong gravity regimes. Different observational techniques using either gravitational waves or  photons in practically every segment of the electromagnetic spectrum  allow us to study black hole mergers, formation, shadows, accretion disk processes, gravitational lensing, star disruption process due to the presence of a black holes and so on \cite{Akiyama1,Akiyama2,AbbottBH}. These tests often put very strong constraints on any modification of general relativity.


While general relativity has been successfully tested on solar system scales, it does appear to run into problems at galactic  and larger scales where we are forced to postulate the existence of exotic components like dark mater and dark energy. We also know that general relativity is incomplete since it is a classical theory and cannot describe space-time at the microscopic scales. The best developed candidate that might provide us with quantum description of gravity is string theory. However, making any testable prediction of the string theory proved to be notoriously difficult. Thus, any headway in this direction, albeit not quite conclusive, might be very important for further development of the field.     

In this work, we consider the D-dimensional $\alpha'$- corrected (where $\alpha'$ is the string coupling constant)  black hole found by Callan, Myers and Perry (CMP) \cite{callan}. Originally, the CMP solution was found perturbatively, with only leading order stringy corrections. Nonetheless such a solution is important since it can give us many important insights about Hawking evaporation and related information loss paradox.  This black hole solution has attracted a great deal of attention from the theoretical point of view, and was even recently used to study some potentially observational effects like the quasinormal modes and black hole shadows in the eiokonal limit \cite{Moura1,Moura2}. To make the original CMP solution relevant for our observed macroscopic 4D world, we have to find a way to obtain a corresponding effective 4D solution. For that purpose, we are inspired by the recent formulation of the Gauss-Bonnet gravity in 4D, which is formulated by rescaling of the coupling parameter $\alpha\to \alpha/D-4$ \cite{Glavan:2019inb}. In the present work, we aim to rescale the parameter $\lambda$ (which is directly proportional to $\alpha'$) as $\lambda \to \lambda/D-4$ to find an effective solution in 4D. 

 In general, 4D Gauss-Bonnet gravity admits many exact spherically symmetric black hole solutions, which attracted a lot of interest (see Refs. \cite{Konoplya:2020qqh,Jusufi:2020qyw,Jusufi:2020yus,Gurses:2020ofy,Gurses:2020rxb,Ai:2020peo,Arrechea:2020gjw,Fernandes:2020rpa,Clifton:2020xhc,Fernandes:2021ysi}). It is important to note that a regularized  4D theory at the level of action was shown to exist, as a special case of the 4D Horndeski theory \cite{Fernandes:2020nbq,Hennigar:2020lsl}. For the static and spherically symmetric black hole solutions, it was shown that the regularized solution (obtained by rescaling the coupling constant) coincides with the original solution obtained by \cite{Glavan:2019inb}. However, to go beyond spherical symmetry, one must use the regularized field equations coming from the regularized action and not from the rescaling $\alpha\to \alpha/D-4$. 

This paper is organized as follows. In Sec. 2, we review the D-dimensional CMP black hole solution. In Sec. 3, we rescale the coupling constant to obtain the 4D SCBH and analyze its properties. We discuss the black hole thermodynamics, Hawking radiation, the ADM mass etc. In Sec. 3, we use the metric we obtained to analyze motion of the S2 star in order to constrain the geometry of a 4D SCBH. In Sec. 4, we study the light deflection angle and the Einstein rings in the weak gravity regime. In Sec. 5, we elaborate the shadow images and the electromagnetic radiation of infalling gas model due to the string corrections. Finally, we discuss our results in Sec. 6. 

\begin{figure*}
\centering
\includegraphics[width=3.45 in]{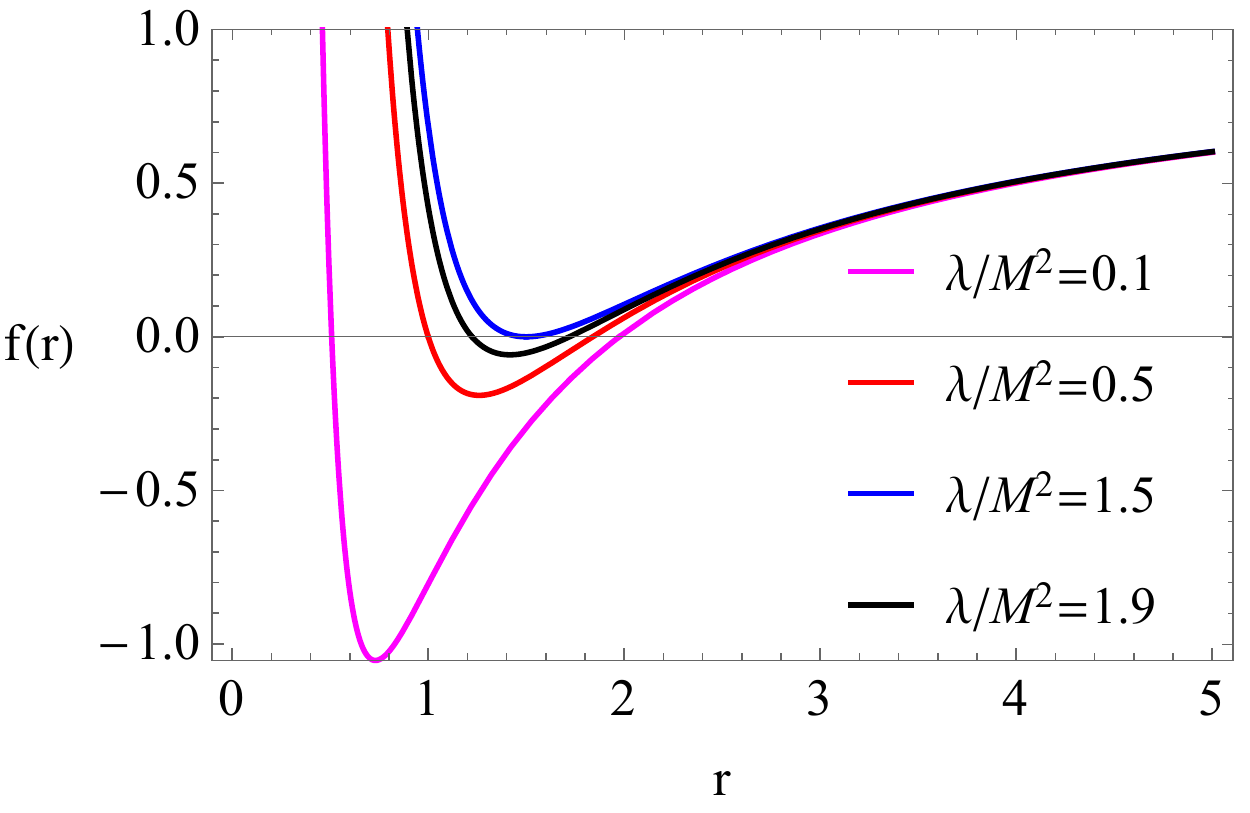}
\includegraphics[width=3.2 in]{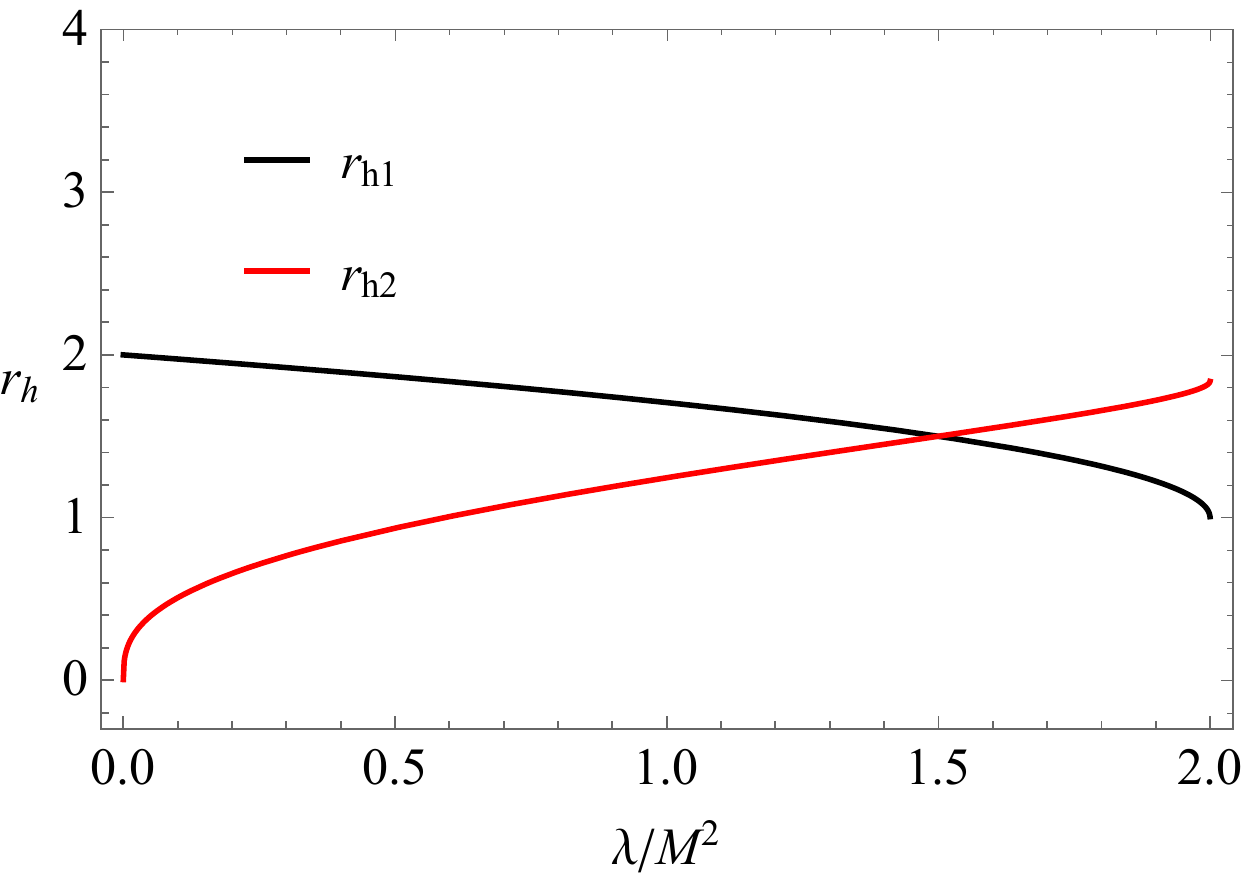}
\caption{Left panel: The plot shows the metric function for different values of $\lambda$. Right panel: The plot shows the two horizons by varying $\lambda$ while keeping $M=1$. The two horizons coincide at $\lambda_c$. \label{fig1} }
\end{figure*}

\section{D-Dimensional String-Corrected Black Hole \label{sect2}}
We start with the expression for the $D$--dimensional effective action with purely gravitational corrections given by \cite{callan,Moura1,Moura2}
\begin{eqnarray}\notag \label{action}
 \mathcal{S}&=&\frac{1}{16 \pi G} \int \sqrt{-g} \Big( R -
\frac{4}{D-2} \left( \d^\mu \phi \right) \left(\d_\mu \phi\right) \\
&+&
\mbox{e}^{-\frac{4}{D-2} \phi} \frac{\lambda'}{2}\
R^{\mu\nu\rho\sigma} R_{\mu\nu\rho\sigma} \Big) \mbox{d}^D x .
\end{eqnarray} 
Note that the above effective action encodes the bosonic and heterotic string theories, to the first order in $\a$, with $\lambda' = \frac{\a}{2}, \frac{\a}{4}$, respectively. We are interested in a  general static and spherically symmetric solution in $D$ dimensions which can be written as follows
\be \label{schwarz}
ds^2 =f(r)\ dt^2  - f^{-1}(r)\ dr^2 - r^2 d\Omega^2_{D-2}.
\ee
where
\begin{equation}
  d\Omega^2_{D-2}=d{\theta_1}^2+{\sin^2{\theta_1}^2}{d{\theta_2}^2}+ . . . +\prod_{i=1}^{D-3}\sin^2\theta_id{\theta}_{D-2}^2,
\end{equation}
represents a metric on a $(D-2)$-dimensional unit sphere. We will consider only
gravitational terms, i.e. we neglect all the fermionic and gauge fields. However, as was pointed out in \cite{callan,Moura1,Moura2}, the dilaton field does not decouple, as it can be seen from the field equations
(neglecting terms which are quadratic in $\phi$) \cite{callan,Moura1,Moura2}:
\begin{equation}
\nabla^2\phi - \frac{\lambda'}{4}\ \mbox{e}^{\frac{4}{2-D} \phi} \left(
R_{\rho\sigma\lambda\tau} R^{\rho\sigma\lambda\tau} \right) =0,
\end{equation}
\begin{eqnarray}\notag
 &&\lambda'\ \mbox{e}^{\frac{4}{2-D}
\phi} \left( R_{\mu\rho\sigma\tau} {R_{\nu}}^{\rho\sigma\tau} -
\frac{1}{2(D-2)} g_{\mu\nu} R_{\rho\sigma\lambda\tau}
R^{\rho\sigma\lambda\tau} \right)\\
&+& R_{\mu\nu}=0.
\end{eqnarray}

So far only the first order $\a$ corrections to $\phi$ and $g_{\mu\nu}$ have been calculated. In other words, working perturbatively in $\lambda',$ and neglecting $\lambda'^2$ and higher
order terms, the solution for the $D-$dimensional black hole is given by the Callan-Myers-Perry metric in terms of the following relation  \cite{callan,Moura1,Moura2}
\begin{eqnarray} \label{dfr} 
f(r) &=& \left(1 - \frac{R_H^{D-3}}{r^{D-3}}\right) \left(1+ \frac{\lambda'}{R_H^2} \delta f(r) \right), 
\end{eqnarray}
\begin{eqnarray} \label{dfr1} 
\delta f(r) &=& - \frac{(D-3)(D-4)}{2}\ \frac{R^{D-3}_H}{r^{D-3}}\ \frac{1 - \frac{R_H^{D-1}}{r^{D-1}}}{1 - \frac{R^{D-3}_H}{r^{D-3}}}.
\end{eqnarray}

Such a black hole has the following Hawking temperature \cite{callan,Moura1,Moura2}
\be
T = \frac{D-3}{4 \pi R_H} \left( 1 - \frac{\left( D-1 \right) \left( D-4 \right)}{2}\ \frac{\lambda'}{R_H^2} \right). \label{temp}
\ee

Obviously, $R_H$ is the horizon of the black hole. If we neglect the $\a-$corrections, we recover the  Tangherlini solution in the limit $\lambda'=0$, i.e., $f(r)=f_{\lambda'=0}(r)$.  Furthermore, by setting $D=4$, we effectively eliminate $\a-$corrections in 4D. 

\section{The 4D string corrected black hole}

In order to obtain an effective metric in 4D and still keep stringy corrections, one of the possibilities is to rescale the parameter $\lambda$ as follows
\begin{eqnarray}\label{rs}
    \lambda' \rightarrow \frac{\lambda}{D-4},
\end{eqnarray}
and then take the limit $D \to 4$ in Eqs. (\ref{dfr}-\ref{dfr1}). We thus obtain 
\begin{eqnarray}
    f(r)=\left(1-\frac{R_H}{r}\right)\left(1-\frac{\lambda}{2 R_H r }\ \frac{1 - \frac{R_H^{3}}{r^{3}}}{1 - \frac{R_H}{r} } \right).
\end{eqnarray}
This solution can also be written as 
\begin{eqnarray} \label{fr}
    f(r)=1-\frac{R_H}{r}+\frac{\lambda\left( R_H^3-r^3\right)}{2 R_H r^4}.
\end{eqnarray}

In this paper we take $R_H$ and $\lambda$ to be free parameters. 
The original perturbation parameter $\lambda'$, before rescaling in eq. (\ref{rs}), is a small parameter in appropriate units. In the opposite limit, when  $\lambda' \rightarrow \infty$, we have a tensionless limit of string theory, where in general one cannot ignore the higher-order corrections.  To keep the perturbative corrections small in eq. (\ref{dfr}) after the rescaling (at least in the first order of perturbations), we have to require that the parameter $\lambda$ is also small.
In most of the analysis in this paper we will assume that $\lambda$ is small, however we will sometimes relax this condition to extrapolate the results and obtain some interesting findings. 
We should note however that, strictly speaking, we lose the perturbative connection with the original action in (\ref{action}) and simply analyze the metric given by the element in eq. (\ref{fr}) for arbitrary values of parameters. Note that since $\lambda$ is proportional to the string coupling, which on the other hand is related to the string tension, we will assume that $\lambda$ is a positive quantity.

Solving for $f(r)=0$, we find two real solutions for the horizons located at 
\begin{equation}
  r_{h1} = R_H,\,\,\,\,  r_{h2}=\frac{\lambda}{6R_H}+\frac{\lambda^{1/3} Z^{1/3}}{6R_H}+\frac{\lambda (6R_H^2+\lambda)}{6 R_H Z^{1/3}},
\end{equation}
where 
\begin{eqnarray}\notag
    Z&=& 3 \sqrt{3} \sqrt{108R_H^4+28 \lambda R_H^2+3 \lambda^2}\,\, R_H^2+54R_H^4\\
    &+& 9 \lambda R_H^2+\lambda^2.
\end{eqnarray}
 Using this metric function in 4D we can compute the scalars
\begin{eqnarray}\notag
    R_{\mu\nu\alpha\beta}R^{\mu\nu\alpha\beta}&=& \frac{12 R_H^2}{r^6}+\left(\frac{12 r^6 R_H^2-60 R_H^5 r^3}{r^{12} R_H^2}\right)\lambda \\
    &+& \left(\frac{117 R_H^6-30 R_H^3 r^3+3 r^6}{r^{12} R_H^2}\right)\lambda^2,\label{eq4}
\end{eqnarray}
and
\begin{eqnarray}
  R^{\mu\nu}R_{\mu\nu}=\frac{3 R_H^2 \lambda}{r^6}.\label{eq15}
\end{eqnarray}
From these expressions it looks like there is a singularity in the limit of $r \to 0$, which means that perturbative string corrections cannot resolve the problem of the central singularity in this case. In the next section, we shall argue more about the nature of this singularity.  In addition, there will be important effects on the black hole evaporation as we will see later.

\subsection{The ADM mass}
Let us first rewrite the metric (\ref{fr}) in the following form
\begin{eqnarray}
    f(r)=1-\frac{2\left(R_H/2+\lambda/4R_H\right)}{r}+\frac{\lambda R_H^2}{2 r^4}.\label{m16}
\end{eqnarray}
It is natural to ask if we can identify the quantity $R_H/2+\lambda/4R_H$ with the black hole mass? We can see that answer to this question is yes! 
We can compute the ADM mass for the 4D SCBH and show that, in fact, it is a well defined quantity. Since our metric is asymptotically flat, to compute the ADM mass, we can apply the relation used in  \cite{Shaikh:2018kfv},
\begin{equation}\label{for}
    M_{ADM}=\lim_{r\to \infty} \frac{1}{2}\left[-r^2 \chi'+r(\psi -\chi) \right],
\end{equation}
where we have identified
\begin{equation}
    \psi(r)=\frac{1}{f(r)},\quad \text{and} \quad \chi(r)=1.
\end{equation}

After simple calculations we can see that the ADM mass for our black hole given by
\begin{equation}\label{ADM}
    M_{ADM}=\frac{R_H}{2}+\frac{\lambda}{4R_H}.
\end{equation}
The last equation represents the mass of the 4D SCBH measured by an observer which is located at the
asymptotic spatial infinity.  We can also verify that the AMD mass coincides with the Misner--Sharp mass defined as 
\begin{eqnarray}
    M=\lim_{r \to \infty} \frac{r}{2}\left(1-g^{ab} (\partial_a r)(\partial_b r)   \right).
\end{eqnarray}
Taking the limit of the last equation, as expected, we obtain equation (\ref{ADM}). We can now simplify the notation and refer to the ADM mass simply as $M=M_{ADM}$. Thus we can write the metric (\ref{m16}) in terms of three quantities: the black hole mass, and parameters $\lambda$ and $R_H$ as follows 
\begin{eqnarray}
    f(r)=1-\frac{2M}{r}+\frac{\lambda R_H^2}{2 r^4}.\label{m21}
\end{eqnarray}
\begin{figure*}
\centering
    \includegraphics[width=3.23 in]{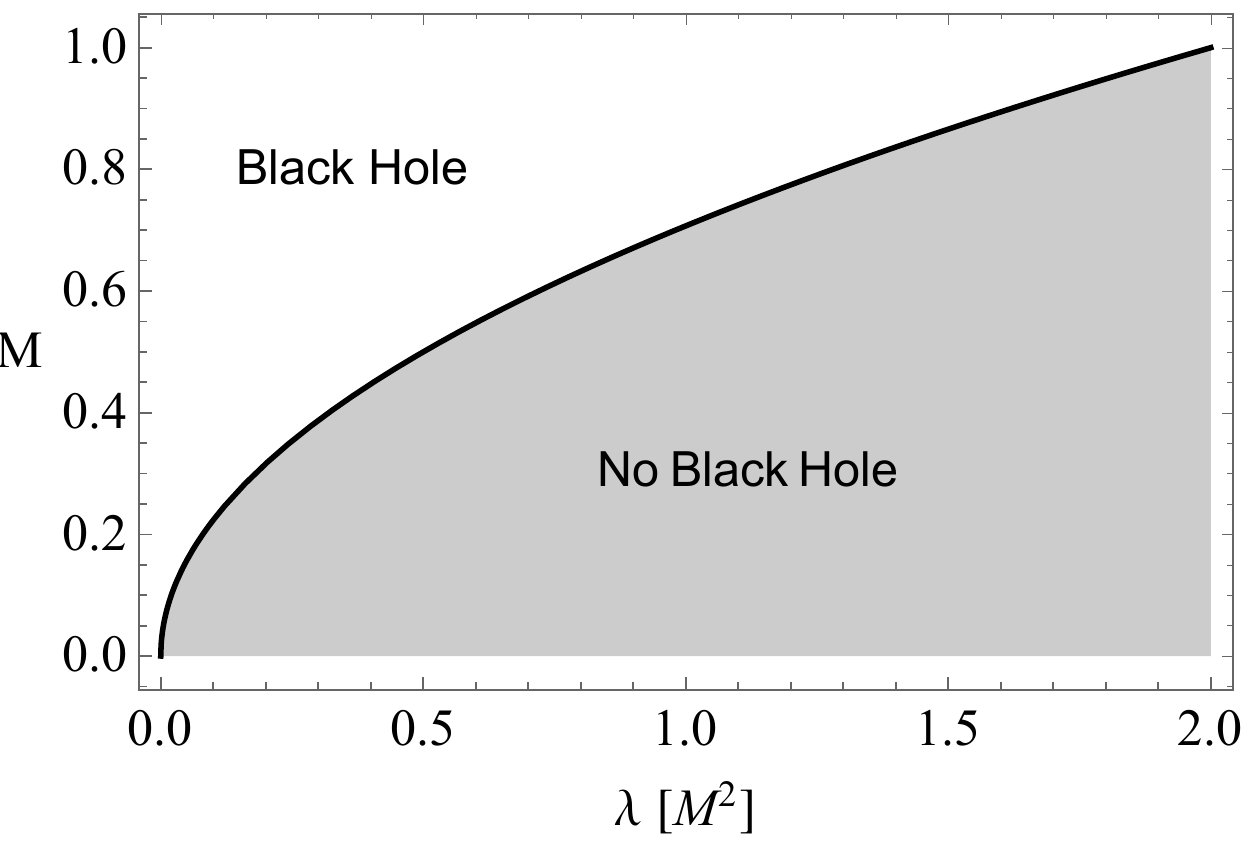}
      \includegraphics[width=3.5 in]{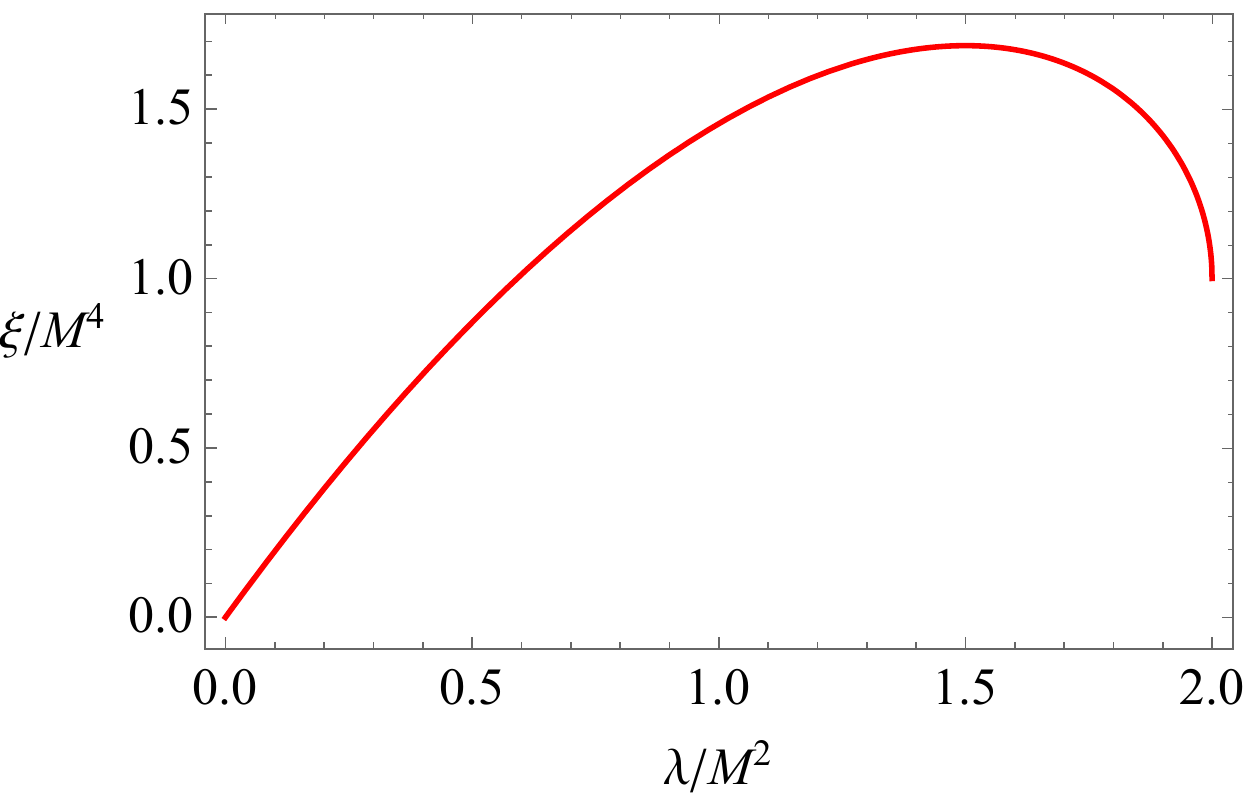}
\caption{Left panel: Parametric plot between the mass of the black hole and the parameter $\lambda$. Right panel: Parametric plot between $\xi$ and the parameter $\lambda$. We have set $M=1$. \label{fig2}}
\end{figure*}

\begin{figure*}
\centering
    \includegraphics[width=3.23 in]{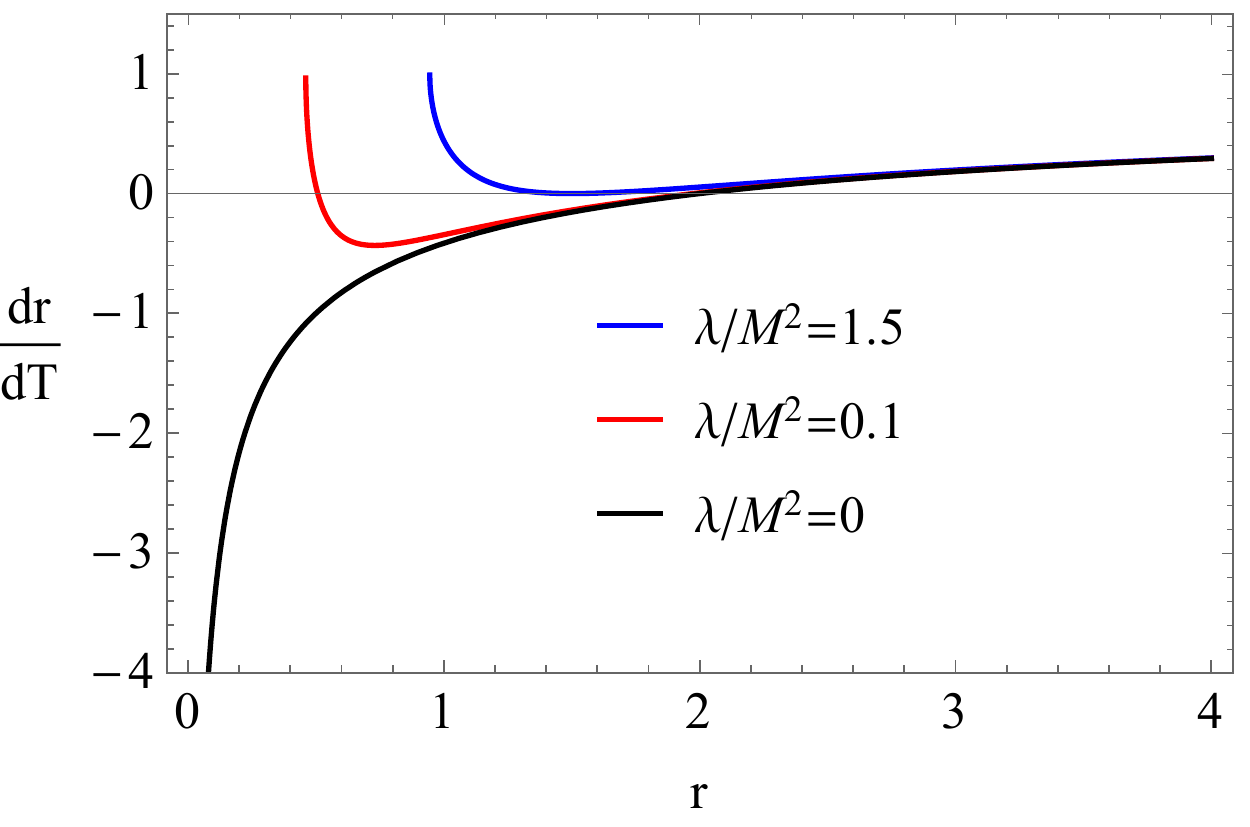}
      \includegraphics[width=3.23 in]{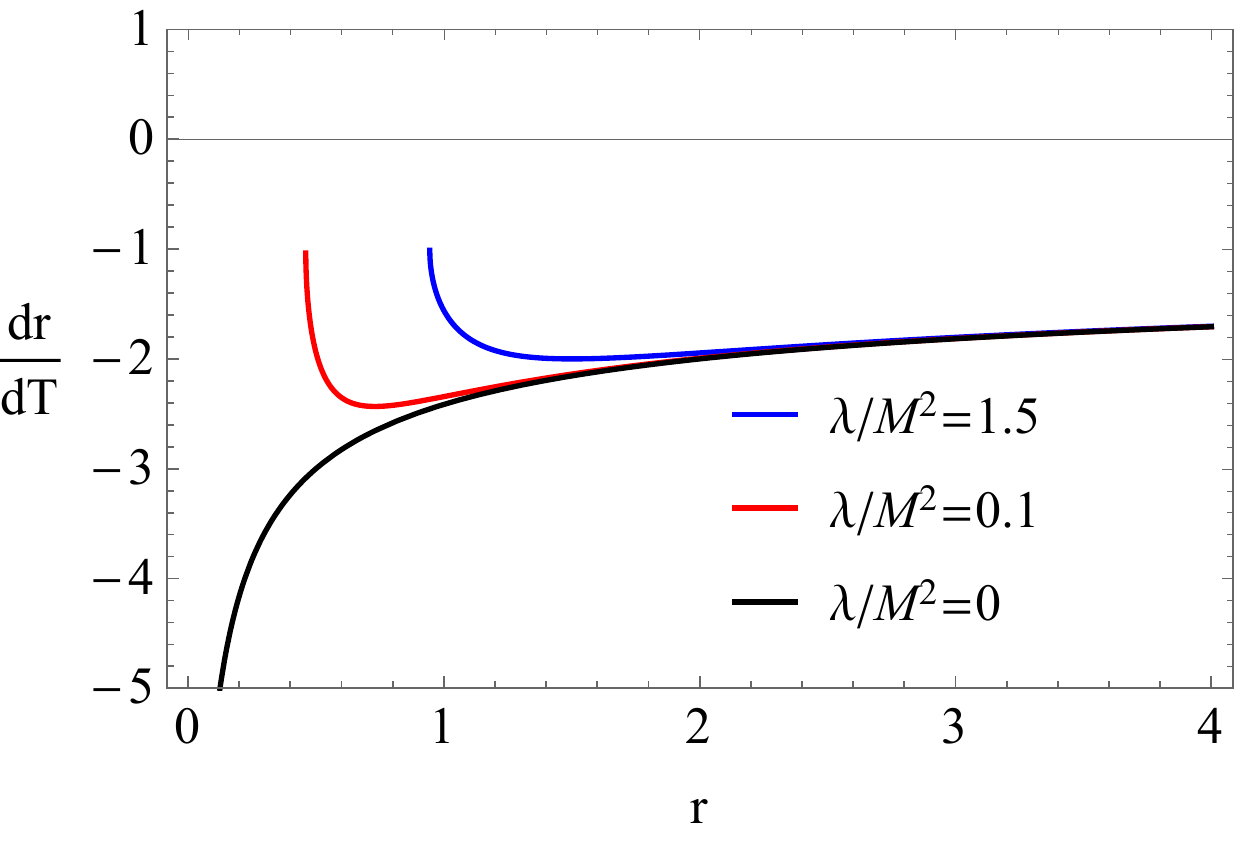}
\caption{Left panel: The velocity of light observed in a freely falling frame for light moving inward. Right panel: The velocity of light observed in a freely falling frame for light moving outward. We have set $M=1$. \label{fig3}}
\end{figure*}

We can solve the expression of the ADM mass (\ref{ADM}) for $R_H$ to obtain two solutions 
\begin{eqnarray} 
    R_H=M\pm \frac{\sqrt{4M^2-2\lambda}}{2}.\label{RH}
\end{eqnarray}
 In the limit of $\lambda \to 0$, we would like to obtain the Schwarzschild BH, hence we accept  only the positive branch as a physical solution. This also means that $R_H$ decreases with due to the stringy corrections.  In addition, the last equation requires $4M^2-2\lambda\geq 0$ for a horizon to exist, which gives the following constraint $\lambda/M^2 \leq 2$. In Fig. \ref{fig1} we show the metric function $f(r)$ and the horizons for different values of $\lambda$. In Fig. \ref{fig2} (left panel) we show the parametric region of $M$ and $\lambda$.  Since $\lambda$ and $R_H$ are just parameters we can introduce a new quantity, $\xi$,  and write metric function (\ref{m21}) as follows 
\begin{eqnarray}
    f(r)=1-\frac{2M}{r}+\frac{\xi}{r^4},\label{m23}
\end{eqnarray}
where
\begin{eqnarray}
    \xi = \frac{\lambda R_H^2}{2}=\frac{\lambda M^2}{2}\left( 1+\sqrt{1-\frac{ \lambda}{2 M^2}}\right)^2,
\end{eqnarray}
provided $\lambda/M^2 \leq 2$. It is very interesting to note that the Kretschmann scalar and the Ricci scalar  (see, Eqs. (\ref{eq4}-\ref{eq15})) goes to infinity as $r \to 0$. However, in contrast with the  Schwarzschild black hole, we can see here that due to the string corrections, light rays (and particles) never reach  the singularity.  To see this fact, let us consider a reference frame of an observer falling from rest towards the black hole. For this purpose we use the Painlev\'e-Gullstrand coordinates. Introducing $dT=dt-h(r) dr$, in the equatorial plane ($\theta=\pi/2$), after fixing $h(r)=\sqrt{1-f(r)}/f(r)$, we obtain the metric
\begin{eqnarray}
    ds^2=f(r)dT^2-2 \sqrt{\frac{2M}{r}-\frac{\xi}{r^4}}dTdr-dr^2,
\end{eqnarray}
which is regular at the black hole horizon. If we solve for radial light geodesics $ds^2=0$, one can find two solutions
\begin{eqnarray}
    \frac{dr}{dT}=-\sqrt{\frac{2M}{r}-\frac{\xi}{r^4}} \pm 1,
\end{eqnarray}
where the positive sign represents light moving in the outward direction, while the negative sign
represents light moving in the inward direction, respectively. In Fig. \ref{fig3}, we see that for the Schwarzschild black hole (black curve), the velocities of both the inward and outward light rays go to negative infinity as $r \to 0$, this physically means that anything that enters the event horizon of the black hole reaches the singularity at the center. In contrast with this behavior, when we inlcude the string corrections, we find that the velocity reaches a minimum
value at some $r$ but then increases for smaller values of $r$. The velocity
of an inward light ray reaches $-1$, and the velocity of an outward light ray reaches $1$ (see fig. \ref{fig3}). Below this value, $dr/dT$ is imaginary, which means that nothing can reach the center, and the
observed curvature scalars remain finite. Such a singularity is known as a timelike singularity, which is different from the Schwarzschild black hole which harbors a spacelike singularity. A similar effect was observed in Ref. \cite{Ali:2015tva} where the quantum corrected black hole was investigated. 

\begin{figure*}
\centering
 \includegraphics[width=3.3 in]{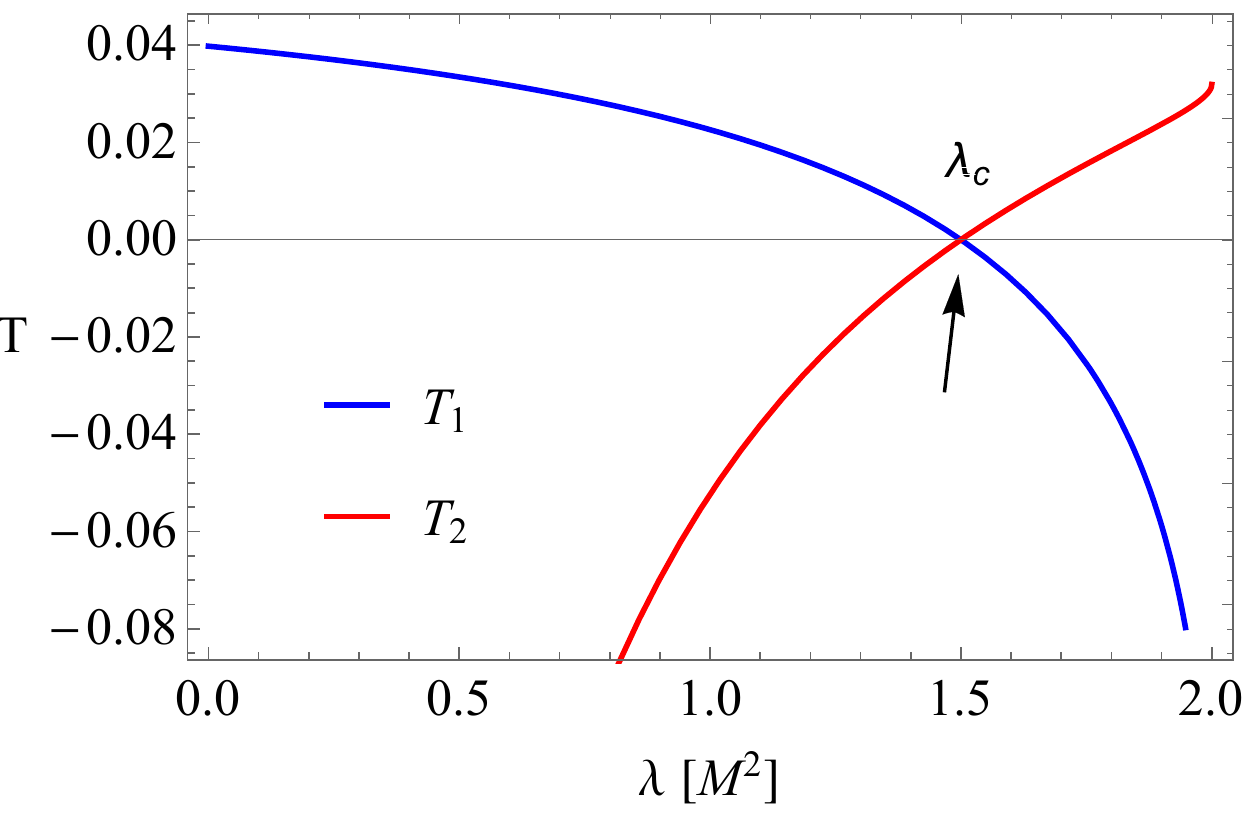}
   \includegraphics[width=3.3 in]{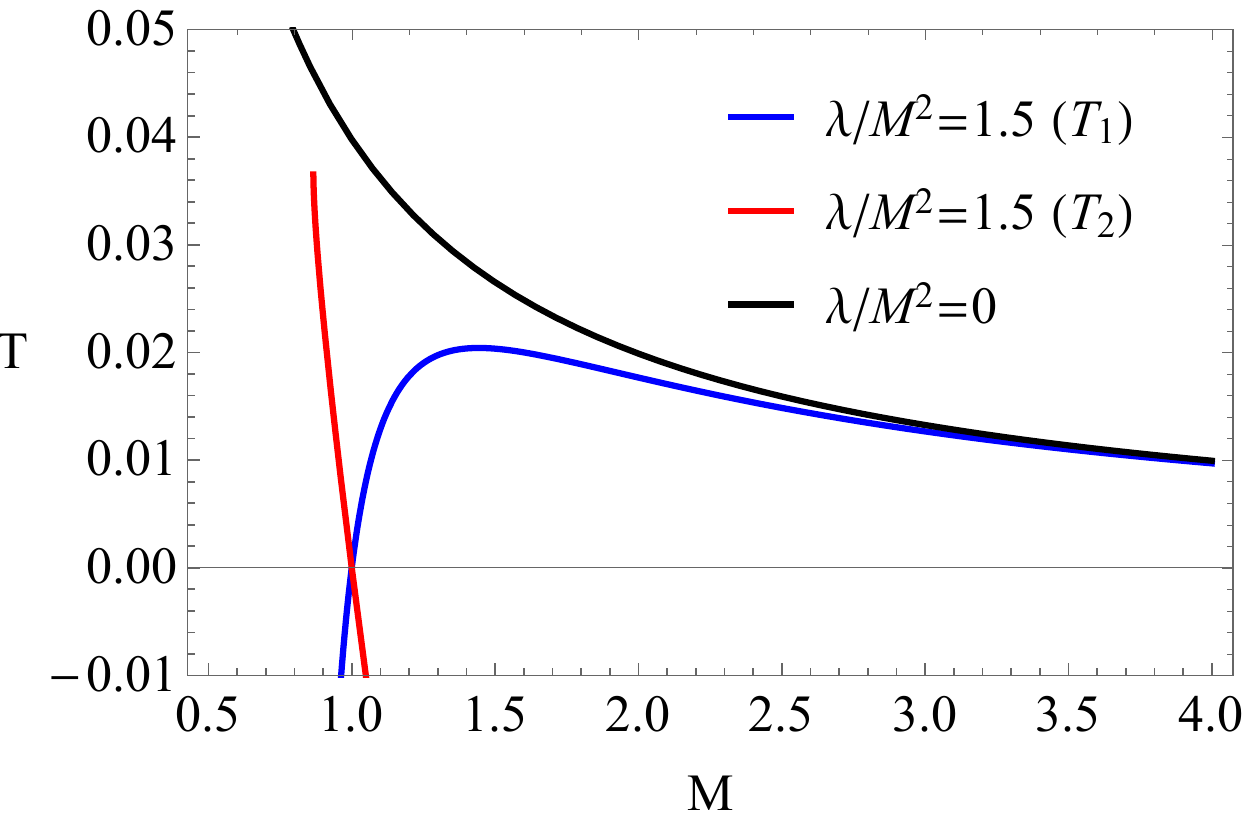}
\caption{Left panel:The plot showing  Hawking temperature of the 4D SCBH as a function of the coupling parameter $\lambda$ for a fixed mass of the black hole $M=1$. Right panel:The plot showing  Hawking temperature of the 4D SCBH as a function of black hole mass for a fixed values of the coupling parameter $\lambda$. \label{fig4}}
\end{figure*}

If $\lambda/M^2 > 2$,  the horizon  becomes complex and hence physically absent. This would be usually interpreted as a naked singularity, however, the curvature invariants also become complex, so interpretation of the solution in that regime is not clear. It might happen that, in this domain of $\lambda$, a gravitational phase transition happens and the transition from the black hole to a regular objects occurs.  

Note that $\xi$ is measured in units of $M^4$. Moreover, one can check analytically, by taking the first derivative of $\xi$ with respect to $\lambda$, that the maximum value of $\xi_{max}/M^4=27/16 \simeq 1.68$ is obtained at the same critical value $\lambda_c=1.5$. We can see this in Fig. \ref{fig2} (right panel). It is important to specify whether we work with metric (\ref{fr}) or (\ref{m23}), of course, the physics remains the same but the range of parameter changes.  Furthermore if we consider an expansion around $\lambda$, from metric (\ref{m23}) we find
\begin{eqnarray}
    f(r)=1-\frac{2M}{r}+\frac{2 M^2 \lambda}{r^4}+...\label{m27}
\end{eqnarray}

If we compare our solution to the recent 4D Gauss-Bonnet solution 
\begin{eqnarray}
    f(r)=1+\frac{r^2}{2 \alpha}\left(1-\sqrt{1-\frac{8M \alpha}{r^3}}  \right),
\end{eqnarray}
and expand the last equation to the leading order term we find 
\begin{equation}
    f(r)=1-\frac{2M}{r}+\frac{4M^2 \alpha}{r^4}+....
\end{equation}
which coincides with (\ref{m27}) if we rescale $\lambda \to 2 \alpha$. This is an indication of the perturbative equivalence between the 4D Gauss-Bonnet solution and our 4D SCBH. 


\subsection{Hawking radiation}

 Let us write the corrected Hawking temperature using the metric (\ref{fr}) associated with the horizon $r_{h1}$ as
\be
T_1 =\frac{f'(r)}{4 \pi}|_{r=r_{h1}}= \frac{1}{4 \pi R_H} \left( 1 - \frac{3}{2}\ \frac{\lambda}{ R_H^2} \right) \label{temp1} .
\ee
It can be seen that for the critical value of the parameter
\begin{eqnarray}
    \lambda_c=\frac{2\,R_H^2}{3}
\end{eqnarray}
the Hawking temperature becomes zero, meaning that the black hole will stop radiating. In that case, small black holes do not disappear and remain as remnants.  If now combine the last condition with Eq. (\ref{RH}) we obtain
\begin{eqnarray}
    \lambda_c=\frac{3}{2}M^2.
\end{eqnarray}
This equations really shows that at this critical value of $\lambda=\lambda_c$ a black hole with mass $M$ will have zero Hawking temperature and, in general, the critical value of $\lambda$ depends on the black hole mass. We can see what exactly happens in Fig. \ref{fig4}. If we fix for simplicity the black hole mass to unity [$M=1]$, in the interval $\lambda \in (1.5, 2)$, the temperature $T_1$ becomes negative, which is not physically acceptable.  But as we will explain below, this is a consequence of the fact that for $\lambda>\lambda_c$, the horizons switch places, and the outer horizon will be  located at $r_{h2}$. So in this region, Hawking temperature should be associated with the horizon $r_{h2}$, that is 
\be
T_2 =\frac{f'(r)}{4 \pi}|_{r=r_{h2}}= \frac{1}{4 \pi} \left(\frac{R_H}{r^2}-\frac{\lambda(4 R_H^3-r^3)}{2  r^5 R_H}\right)|_{r=r_{h2}}. \label{temp2}
\ee

At the same critical value $\lambda_c=1.5$ where the black hole stops evaporating, the two horizons coincide, i.e., $r_{h1}=r_{h2}=1.5$ (see Fig. \ref{fig1}, right panel). In the interval $\lambda \in (0, 1.5)$, the outer horizon is $r_{h1}$, meaning that $r_{h1}>r_{h2}$, at the critical value $\lambda_c$ the two horizons coincide, while at the interval $\lambda \in (1.5, 2)$, the outer horizon is now $r_{h2}$, meaning that $r_{h2}>r_{h1}$.  Therefore the Hawking temperature in the total interval is only the positive contribution obtained from the interval
\be
T= \begin{cases}
T_1, &\lambda \in (0, 1.5)\\
T_2, & \lambda \in (1.5, 2)
\end{cases}
\ee
 We can see this fact also from the Fig. \ref{fig4}. If we take $\lambda/M^2=1.5$, we see that as the black hole mass decreases the Hawking radiation increases and attains a maximum value when $(\partial T_{h1}/\partial r_{h1})=0$, we get the maximal value of the Hawking temperature at $M=1.443375673$. Now if we further decrease the mass, we see that at $M=1$, the Hawking temperature is exactly zero.  On the other hand, if we have a black hole with mass in the range $M<1$, then there is a nonzero Hawking radiation coming from the horizon $r_{h2}$, with a Hawking temperature that increases and reaches a finite value at some minimal possible mass.  We can get the minimal black hole mass using the condition $\lambda/M^2 \leq 2$, in our case [setting $\lambda=3/2$] we obtain $M_{min}=\sqrt{\lambda/2}=\sqrt{3}/2$. This can be also seen in Fig. \ref{fig2} (left panel) or in Fig. \ref{fig4} (right panel). Beyond this minimal mass the black hole  cannot exist, so we end up with some regular object or naked singularity. 
 
 To find the entropy of the black hole we can use the first law of thermodynamics $\frac{\partial S}{\partial M}=\frac{1}{T}$ along with the expression for the temperature. Again we will have two intervals for $\lambda$. The black hole horizon $r_{h1}$ contributes to the entropy in the domain $0<\lambda/M^2<1.5$. It is convenient to express the entropy in terms of the black hole horizon $r_{h1}=R_H$. For the entropy in this interval we find
 \begin{equation}
     S=\int \frac{1}{T_1} \frac{\partial M}{\partial R_H} dR_H=\pi R_H^2+\pi \lambda \ln\left(2 R_H^2-3 \lambda  \right)+\mathcal{C},
 \end{equation}
 where $\mathcal{C}$ is some constant of integration. Note that the black hole entropy consists of two terms; the first term is the standard Bekenstein-Hawking area law followed by the logarithmic string correction term. We can also express this entropy in terms of black hole mass by using eq. (\ref{ADM}).
 
On the other hand, inside the interval $1.5<\lambda/M^2<2$, due to the contribution from the horizon $r_{h2}$, it is not possible to find an analytical closed form for the entropy. However, one can evaluate it numerically.

The thermodynamical stability of the black hole can be found by using the heat capacity $C_h$. The stability of the black hole is related to the sign of the heat capacity. In particular, when $C_h>0$ the black hole is stable, while for $C_h <0$, the black hole is unstable. The heat capacity of the black hole is given by 
\begin{equation}
C_h=\frac{\partial M_h}{\partial T_h}=\frac{\partial M_h}{\partial r_h}\frac{\partial r_h}{\partial T_h}.
\end{equation}
\begin{figure}
\centering
 \includegraphics[width=3.4 in]{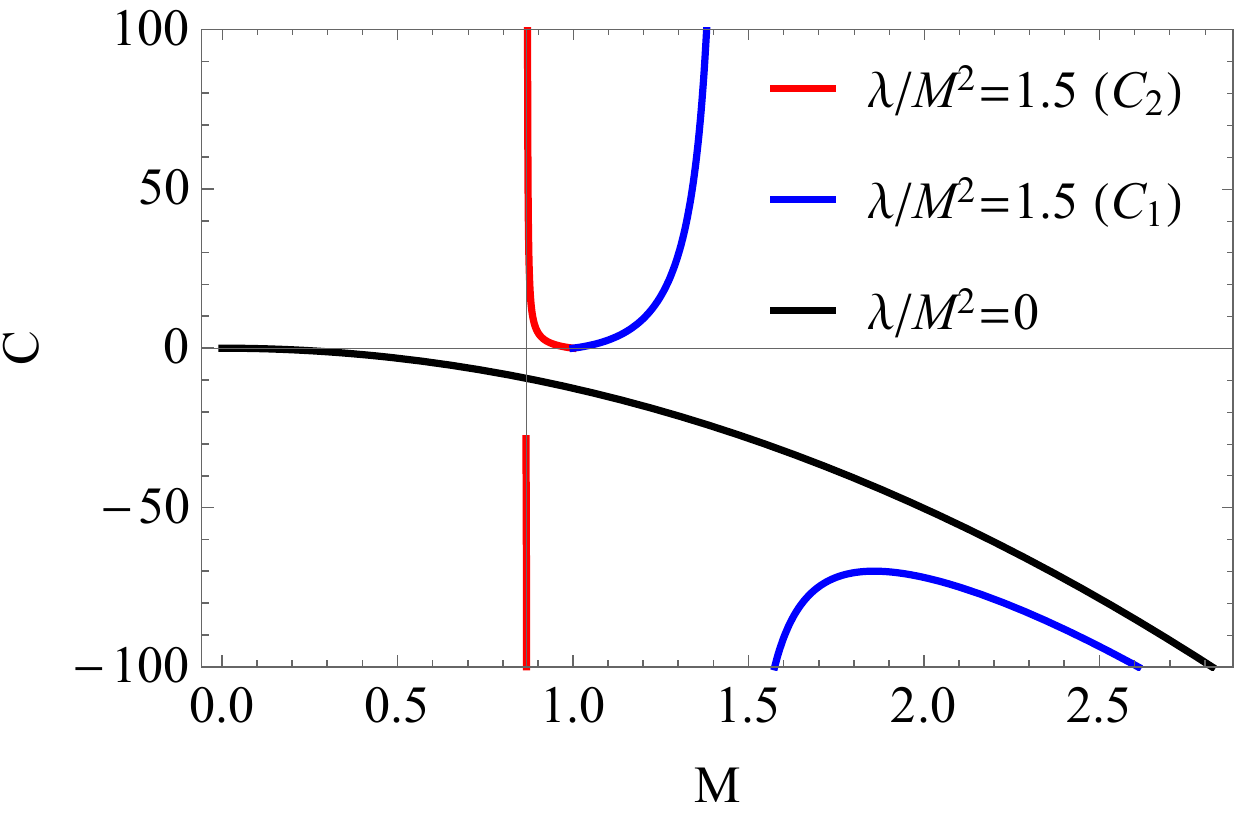}
\caption{The plot showing the heat capacity of the black hole as a function of the black hole mass for a fixed values of the coupling parameter $\lambda$. For $C_h <0$, the black hole is unstable. \label{fig5}}
\end{figure}

In Fig. \ref{fig5} we plot the heat capacity of the black hole as a function of $M$ for a fixed parameter $\lambda/M^2=1.5$. We can see that the total heat capacity $C_h$ of the black hole is obtained from the contribution of the horizon $r_{h1}$ noted as $C_1$ (blue line) in the interval $M\geq 1$, and the contribution of $C_2$ (red line) in the interval $M<1$. In particular one can see that  $C_1$ exhibits discontinuity and diverges at some critical points  $r=r_c$, which can be linked to the second order phase transition.  At the point $r=r_c$, there is a flip of sign in the heat capacity where the Hawking temperature attains a maximum value with $(\partial T_h/\partial r_h)=0$. In the interval $M<1$, we see that $C_2$ is positive and there is a discontinuity at the minimal mass $M_{min}=\sqrt{3}/2$. Beyond this point, the object is not a black hole, instead it can be some regular object or a naked singularity.  This means that, the black hole is thermodynamically stable for $r_h<r_c$ whereas it is thermodynamically unstable for $r_h>r_c$. We also see that in general, for large $M$, we have negative heat capacity $C_h<0$ hence the black hole is unstable. For a fixed value $\lambda/M^2=1.5$, the heat capacity is exactly zero at $M=1$, as expected, since the Hawking temperature becomes zero.  It is also interesting to see that for large values of $M$ the heat capacity gets closer and closer to the Schwarzschild black hole heat capacity, and it is also negative i.e., $C_h<0$.  We note that there is no phase transition for the Schwarzschild black hole as can be seen from Fig. \ref{fig5} (black line). 

\subsection{Can $\lambda$ be a large parameter?}

As we mentioned earlier, the perturbative parameter $\lambda$ should be small in appropriate units.  If that is true, then there is a great level of similarity between our 4D SCBH and the generalized uncertainty principle (GUP) corrected BH. This can be seen in the following. In the limit $\lambda \to 0$, we see from Eq. (\ref{ADM}) that the ADM  mass gives $M_{ADM}=R_H$. This suggests that we can write 
\begin{eqnarray}
    R_H=2M_0,
\end{eqnarray}
where $M_0$ can be interpreted as the bare mass of the black hole. The ADM mass can then be written as 
\begin{eqnarray} \label{gupc}
    M_{ADM}=M_0\left(1+\frac{\lambda/4}{2 M_0^2}\right).
\end{eqnarray}
The second term on the right hand side can be interpreted as the quantum mechanical hair due to the GUP. We start from the GUP equation in the form 
\begin{eqnarray}
    \Delta x = \frac{1}{\Delta p}+\frac{\alpha_{GUP}\,l^2_{\text{Pl}}}{\hbar}\Delta p,
\end{eqnarray}
where $\alpha_{GUP}$ is the GUP parameter, $l_{\text{Pl}}=\sqrt{\hbar G/c^3} \sim 10^{-33}$ cm is the Planck length. We set $\Delta x \to R$, $\Delta p \to c M$, where $M_0$ is the mass forming an event horizon if it falls within its own Schwarzschild radius $R_S=2GM_0/c^2$ [here we have restored the constants $c$, $\hbar$ and $G$]. Using the $\beta$ (GUP corrected constant) formalism, it was shown in Ref. \cite{Carr1} that GUP has an important effect on the size of the black hole as follows
\begin{eqnarray}
    R_S'=\frac{2 G M_0}{c^2}\left(1+\frac{\beta}{2}\left(\frac{M_{\text{Pl}}}{M_0}\right)^2\right)\label{eq40}
\end{eqnarray}
where $M_{\text{Pl}}=\sqrt{\hbar c/G} \sim 10^{-5}$ g. With appropriate units and the scaling 
\begin{equation}
    \beta \to \frac{\lambda}{4},
\end{equation}
it can be easily seen that Eq. (\ref{eq40}) suggests a modified GUP ADM mass with $R_S'=2M_{ADM}$. As a result we can  write 
\begin{eqnarray}
       M_{ADM}=\frac{ G M_0}{c^2}\left(1+\frac{\beta}{2}\left(\frac{M_{\text{Pl}}}{M_0}\right)^2\right).
\end{eqnarray}
The last equation has been obtained in Refs. \cite{Carr1,Carr2}, but it coincides with our result for the ADM mass of the string corrected black hole given by Eq. (\ref{gupc}). This indicates that the string corrections given by (a small) $\lambda$ can be linked to the the GUP corrections $\beta$ of black holes. 

However, $\lambda$ might not necessarily be a small parameter (in appropriate units). 
The rescaling given by Eq. (\ref{rs}), $\lambda' \rightarrow \lambda/(D-4)$, applied in Eqs.(6-7) is fine at the level of the solution for the metric (the terms $D-4$ cancel out), but it is not obviously justified at the level of the action in eq.~(\ref{action}) where singularity will appear in the limit of $D\rightarrow 4$.  However, we can still analyze the metric given by the function in eq. (\ref{fr}), having in mind that we lost perturbative connection with the stringy action in eq. (\ref{action}).
Thus, $\lambda$ might not only encode small stringy corrections to the BH solutions (e.g. like GUP inspired corrections in eq.~(\ref{gupc})), but also large corrections that might completely change physical interpretation of the solution (e.g. like for the critical value $\lambda_c$ in the previous section).    
As it was the case with 4D Gauss-Bonnet gravity, one expects a more general theory with a well defined action under the rescaling in (\ref{rs}). Such a regularized version of the action in eq.~(\ref{action}) is yet to be found, and is outside the scope of the present work. However, it is encouraging that solutions obtained by simple rescaling $\alpha' \rightarrow \alpha'/(D-4)$ in 4D Gauss-Bonnet gravity coincide with solutions following from a well defined regularized action, so we expect that something similar is happening here.  

Incidentally, it was argued in \cite{Mathur:2005zp} that the string tension necessarily becomes very low in Planck units when applied to an object like a black hole, which consists of many individual fundamental quanta. In that case strings become elongated and ``floppy" reaching the size of the black hole horizon, which gives the so-called fuzzball structure to the black hole. In the light of our analysis here, the metric given in terms of the ADM mass of the system (which is the total mass measured by a distant observer located far away from the black hole) is given by the function in eq. (\ref{m23}). This indicates that the string corrections might be hard to detect even if $\lambda$ is large (and thus tension is low), since the corrections decrease as $r^{-4}$.

\section{The S2 star orbit \label{sect3}}

Let us first briefly review the basic calculations needed to obtain the orbit of S2 star assuming that the geometry is described by the 4D QCBH with the metric function (\ref{fr}).  We start with the general D dimensional spherically symmetric spacetime. Due to the spherical symmetry, one can assume that the orbit lies in the equatorial hyperplane  
\begin{equation}
\theta_i (i=1, ...., D-3)=\pi/2, \,\,\, \text{and}\,\,\,\, \theta_{D-2}=\phi, 
\end{equation}

Next, we use the coordinate transformation from spherical Schwarzschild coordinates to Cartesian coordinates given by 
\begin{eqnarray}\notag
    x_{1}&=&  r \sin\theta_1 \hdots \sin\theta_{D-3} \sin\theta_{D-2}\sin\theta_{D-1}\\\notag
    x_{2}&=&   r \sin\theta_1 \hdots \sin\theta_{D-3} \sin\theta_{D-2}\cos\theta_{D-1}\\\notag
    x_{3}&=&   r \sin\theta_1 \hdots \sin\theta_{D-3} \cos\theta_{D-2}\\\notag
    & \vdots & \\\notag
    x_{D-1}&=&   r \sin\theta_1  \cos\theta_{2}\\\notag
    x_{D}&=&   r \cos\theta_1 . \\
\end{eqnarray}
Using the last two equations, we obtain the polar coordinates
\begin{equation}
   x_3 = r \cos\phi ,\qquad x_1= r \sin\phi,\,\,\ x_2=x_4=\hdots=0.
\end{equation}
If we identify  $x_3=x$, $x_1=y$, and $z=0$ we can obtain the equations of motion. We can work with this general $D$-dimensional case and treat the number of spacetime dimension as a parameter, however, we know that as such large distances the spacetime will be effectively 4D.  For this reason, we shall simplify the problem and use our 4D SCBH. From the Lagrangian it follows that 
\begin{eqnarray}\notag
		2 \mathcal{L} &=&\left(1-\frac{R_H}{r}\right)\left(1-\frac{\lambda}{2 R_H r }\ \frac{1 - \frac{R_H^{3}}{r^{3}}}{1 - \frac{R_H}{r} } \right)\dot{t}^2\\
		&-&\frac{\dot{r}^2}{\left(1-\frac{R_H}{r}\right)\left(1-\frac{\lambda}{2 R_H r }\ \frac{1 - \frac{R_H^{3}}{r^{3}}}{1 - \frac{R_H}{r} }  \right)}-r^2 \dot{\phi}^2
\end{eqnarray}
From the spacetime symmetries we have two constants of motion 
\begin{equation}
    \dot{t}=\frac{E}{\left(1-\frac{R_H}{r}\right)\left(1-\frac{\lambda}{2 R_H r }\ \frac{1 - \frac{R_H^{3}}{r^{3}}}{1 - \frac{R_H}{r} } \right)},\,\,\,\,\,\dot{\phi}=\frac{L}{r^2}
\end{equation}
and
\begin{equation}
     \ddot{r} =\dfrac{1}{2 \  g_{rr}(r)}\left[\dfrac{d g_{00}(r)}{dr} \ \dot{t}^2 + \dfrac{d g_{11}(r)}{dr} \ \dot{r}^2 + \dfrac{d g_{\phi \phi}(r)}{dr}  \dot{\phi}^2\right] .\label{eqn:motionr}
\end{equation}

In terms of Cartesian coordinates, we denote the position of the real orbit as $(x, y, z)$, and velocity components $(v_x, v_y , v_z)$. In our present case, using the transformation from spherical Schwarzschild coordinates to Cartesian coordinates:
\begin{equation}\label{eqn:xyz}
   x= r \cos\phi ,\qquad y= r \sin\phi,\,\,z=0,
\end{equation}
we also find the corresponding three-velocities 
\begin{equation}\label{eqn:vxvyvz}
   v_x = v_r \cos\phi - r v_\phi \sin\phi , \qquad v_y = v_r \sin\phi + r v_\phi \cos\phi.
\end{equation}
We will use this to model the orbit of the S2 star in our geometry and compare it with the observational data to constrain two parameters - the mass $M$ and parameter $\lambda$. To do so, we also need to find the apparent orbit on the plane of the sky by relating the coordinates  $(\mathcal{X}, \mathcal{Y}, \mathcal{Z})$ to the real orbit given by $(x, y, z)$ as follows \cite{13}
\begin{align}\notag
    \mathcal{X} &= x B + y G,\\\notag
     \mathcal{Y} &= x A+ y F,\\
     \mathcal{Z} &= x C + y F,
\end{align}
\begin{figure}
\centering
    \includegraphics[width=3.2 in]{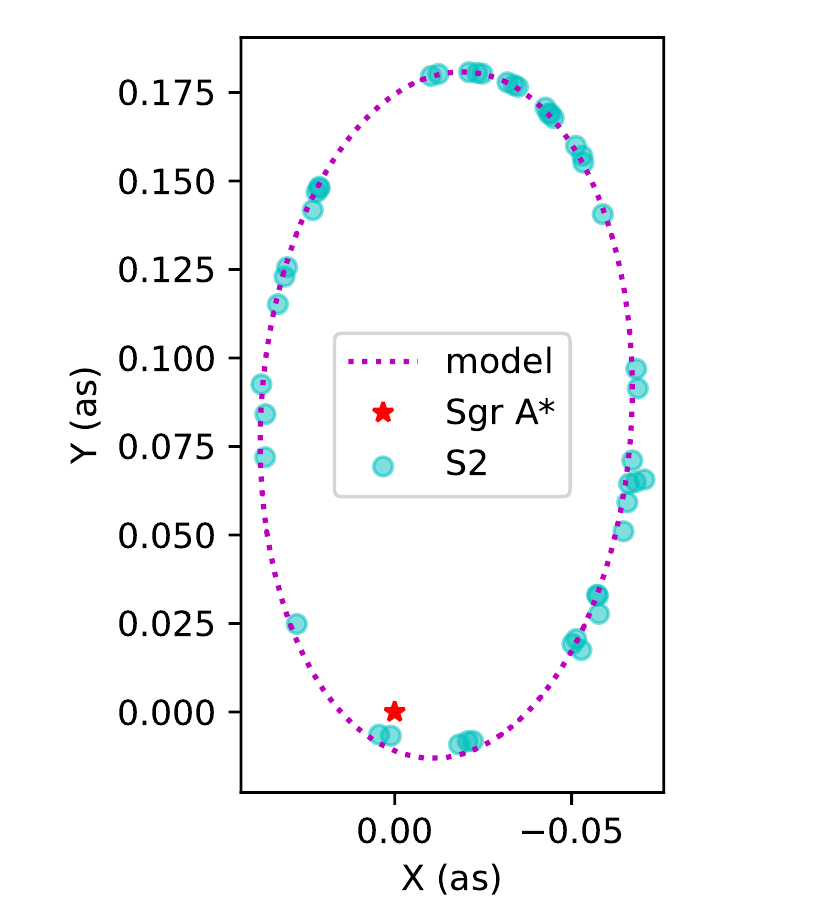}
\caption{The figure shows the orbit fit of the S2 star using the 4D SCBH geometry using the best fit parameters $R_H \sim 1.74[M]$ and  $\lambda \sim 1.10 [M^2]$. The Sgr A$^{\star}$ black hole is assumed to have a mass $M =4.07 \times 10^6 M_{\odot}$. } \label{orbit}
\end{figure}

The corresponding components of the apparent coordinate velocity are follows  \cite{13}
\begin{align}\notag
    \mathcal{V}_X &= v_x B + v_y G,\\\notag
     \mathcal{V}_Y &= v_x A+ v_y F,\\
     \mathcal{V}_Z &= v_x C + v_y F,
\end{align}
where \cite{13}
\begin{eqnarray}
    B &=& \sin\Omega \cos\omega + \cos\Omega \sin\omega \cos i \\
    G &=& -\sin\Omega \sin\omega + \cos\Omega \cos\omega \cos i \\
    A&=& \cos\Omega \cos\omega - \sin\Omega \sin\omega \cos i\\
    F &=& -\cos\Omega \sin\omega - \sin\Omega \cos\omega \cos i\\
    C&=& \sin\omega \sin i\\
    F &=& \cos\omega \sin i.
\end{eqnarray}
where $\omega$, $i$, and $\Omega$ are the argument of pericenter, the inclination between the real orbit and the observation plane, and the ascending node angle, respectively.  Finding an analytical expression for $r(\phi)$ is not possible, therefore we use numerical integration of the equations of motion. To fit our 4D SCBH model, we have to solve the equations of motion numerically using the data presented in \cite{12,13,Jusufi:2021lei,deMartino:2021daj}, and assuming the central mass object \cite{11}. We use the Bayesian theorem according to which the observations O, and the vector containing the parameters of a model, say P, give the posterior probability density $\pi (P | O)$  
\begin{eqnarray}
    \ln \pi(P|O) \propto \ln \mathcal{L}(O|P)+\ln \pi(P),
\end{eqnarray}
where $ \pi(P)$ is the prior probability density of the parameters. 
 The likelihood function is given by 
\begin{eqnarray}\notag
    \ln \mathcal{L}(O|P)&=&-\frac{1}{2}\sum_{i=1}^{N}\left[ \frac{\left(X_{obs,i}-X_{mod,i}\right)^2}{\sigma_{obs,i}^2}\right]\\   
    &-& \frac{1}{2}\sum_{i=1}^{N}\left[\frac{\left(Y_{obs,i}-Y_{mod,i}\right)^2}{\sigma_{obs,i}^2}\right],
\end{eqnarray}
\begin{figure}
\centering
\includegraphics[width=3.6 in]{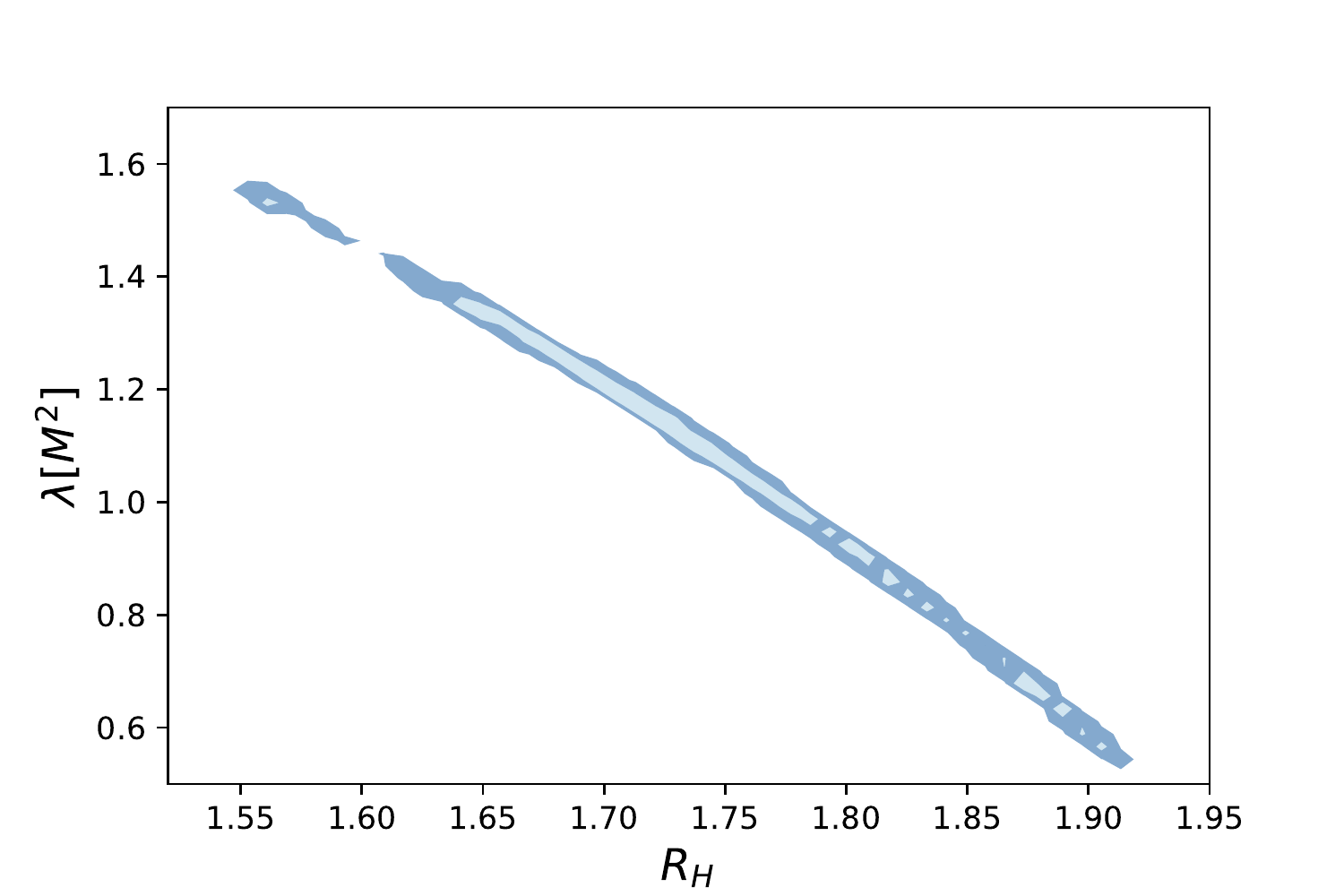}
\caption{The figure shows the parametric plot of $\lambda$ versus $R_H$ with 68\% and 96\% confidence contours.\label{fig7}} \label{contourwh}
\end{figure}

where the two observed and theoretical quantities are noted as $(X_{obs},Y_{obs})$, and $(X_{mod},Y_{mod})$, respectively. To find the best-fitting values we use the Monte-Carlo-Markov Chains
analysis, with the uniform priors $\lambda/M^2 \sim [0, 2]$ along with $R_H/M \sim [0,3]$.  Our analyses show that the best-fitting value for the black hole parameter $R_H$ with $68\%$ confidence level is $R_H \sim 1.74^{+0.12}_{-0.08} [M]$ as shown in Fig. \ref{fig7}, and the orbit fit in Fig. \ref{orbit}. We also find the best fit value for the parameter $\lambda \sim 1.10^{+0.21}_{-0.36} [M^2]$ within $68\%$ confidence level. In SI units, after reinstating $G$ and $c$, for the string coupling parameter we obtain the following value 
\begin{eqnarray}
    \lambda \simeq 1.10\,\times \left(\frac{G M}{c^2}\right)^2 \simeq 4 \times 10^{19}\, \text{m}^2.
\end{eqnarray}
As a comparison, using the precession of the S2 star in orbit around  Sgr A$^{\star}$, the 4D Gauss-Bonnet coupling constant $\alpha_{GB}$ was constrained by an upper bound of $\alpha_{GB}\simeq 10^{25}\,\, \text{m}^2$ (see, \cite{Clifton:2020xhc}). For the mass of the black hole we find 
\begin{equation}
    M=\frac{R_H}{2}+\frac{\lambda}{4 R_H}\simeq 1.028^{+0.078}_{-0.086}\,\,[4.07 \times 10^6 M_{\odot}],
\end{equation}
which is (within the errors) indistinguishable  from the result in the Schwarzschild spacetime reported recently in Refs. \cite{11,12,13,Jusufi:2021lei}. We can also estimate the parameter $\xi=\lambda R_H^2/2 \simeq 1.665 [M^4]$. This indicates that the current precision is insufficient to distinguish the 4D SCBH spacetime from the Schwarzschild spacetime.

\section{Einstein rings in the weak field regime}

To study gravitational lensing, due to the spherical symmetry, we again consider the equatorial hyperplane. 
The photon trajectory (optical metric) is simply found by letting $\mathrm{d}s^2=0$, which gives
\begin{equation}\notag
dt^2=\frac{dr^2}{\left(1 - \frac{R_H^{D-3}}{r^{D-3}}\right)^2\left(1-\frac{\lambda'}{R_H^2} \frac{(D-3)(D-4)}{2}\ \frac{R^{D-3}_H}{r^{D-3}}\ \frac{1 - \frac{R_H^{D-1}}{r^{D-1}}}{1 - \frac{R^{D-3}_H}{r^{D-3}}} \right)^2}
\end{equation}
\begin{equation}
+ \frac{r^2d\phi^2}{\left(1 - \frac{R_H^{D-3}}{r^{D-3}} \right)\left(1-\frac{\lambda'}{R_H^2} \frac{(D-3)(D-4)}{2}\ \frac{R^{D-3}_H}{r^{D-3}}\ \frac{1 - \frac{R_H^{D-1}}{r^{D-1}}}{1 - \frac{R^{D-3}_H}{r^{D-3}}}  \right)}.
\end{equation}
We will closely follow the method in \cite{Gibbons:2008rj} to compute the deflection angle of light. This geometric method is based on the application of the Gauss-Bonnet theorem (GBT) on the optical geometry. We consider a non-singular domain  $\mathcal{A}_{R}$  with boundaries $\partial 
\mathcal{A}_{R}=\gamma_{g^{(op)}}\cup C_{R}$, of an oriented two-dimensional surface $S$ with the optical metric $g^{(op)}$. In terms of the Gaussian optical
curvature ($K$) and the geodesic curvature ($\kappa $), we can write the GBT as follows \cite{Gibbons:2008rj}
\begin{equation}
\iint\limits_{\mathcal{A}_{R}}K\,\mathrm{d}S+\oint\limits_{\partial \mathcal{%
A}_{R}}\kappa \,\mathrm{d}t+\sum_{k}\delta _{k}=2\pi \chi (\mathcal{A}_{R}).
\label{10}
\end{equation}
In which $\mathrm{d}S$ is the optical surface element, $\delta_{k}$ gives the exterior angle at the $k^{th}$ vertex. We can relate the exterior angels with the interior angles (the jump angle at the source $\mathcal{S}$ and observer $\mathcal{O}$, respectively) using $\theta _{\mathcal{O}}=\pi-\delta_{\mathcal{O}}$ and $\theta _{\mathcal{S}}=\pi-\delta_{\mathcal{S}}$, which for a very large radial distance, i.e., $l\equiv R\rightarrow \infty $, satisfies the condition $\theta _{\mathcal{O}}+\theta _{\mathcal{S}}\rightarrow \pi $ (see, \cite{Gibbons:2008rj}]. According to this method, we need to choose the domain of integration to be outside of the light ray in the $(r,\phi)$ optical plane. Moreover this domain can be thought to have the topology of a disc having the Euler characteristic number $\chi (\mathcal{A}_{R})=1$. If we now introduce a smooth curve defined as $\gamma:=\{t\}\to \mathcal{A}_{R}$,  we can find the geodesic curvature in terms of the following definition
\begin{equation}
\kappa =g^{(op)}\,\left( \nabla _{\dot{\gamma}}\dot{\gamma},\ddot{\gamma}%
\right),  
\end{equation}%
 along with the unit speed condition $g^{(op)}(\dot{\gamma},\dot{\gamma})=1$, and $\ddot{\gamma}$ being the unit acceleration vector.  Note that, by definition, the geodesic curvature for the light ray (geodesics) $\gamma_{g^{(op)}}$ vanishes, i.e. $\kappa (\gamma_{g^{(op)}})=0$. One should only compute the contribution to the curve $C_{R}$. Thus, from the GBT we find
\begin{equation}
\lim_{R\rightarrow \infty }\int_{0}^{\pi+\hat{\alpha}}\left[\kappa \frac{d t}{d \phi}\right]_{C_R} d \phi=\pi-\lim_{R\rightarrow \infty }\iint\limits_{\mathcal{A}_{R}}K\,\mathrm{d}S
\end{equation}
The geodesic curvature for the curve $C_{R}$ located at a coordinate distance $R$ from the coordinate system chosen at the black hole center can be calculated via the relation 
\begin{equation}
\kappa (C_{R})=|\nabla _{\dot{C}_{R}}\dot{C}_{R}|.
\end{equation}
With the help of the unit speed condition, one can show that the asymptotically Euclidean condition is satisfied:
\begin{eqnarray}
\lim_{R\rightarrow \infty } \left[\kappa \frac{\mathrm{d}t}{\mathrm{d}\phi}\right]_{C_{R}}=1.
\end{eqnarray}
From the GBT it is not difficult to solve for the deflection angle which gives
\begin{equation}
\hat{\alpha}=-\int\limits_{0}^{\pi }\int\limits_{r=\frac{b}{\sin \phi}%
}^{\infty } K \mathrm{d}S. 
\end{equation}
Since we are interested in leading order terms, in the last equation for the integration domain we have use an equation for the light ray is $r(\phi)=b/\sin \phi $, where $b$ is the impact parameter.  The Gaussian optical curvature takes the form: 
\begin{eqnarray}\notag
     K &\simeq &  - \frac{(D-3)(D-2)}{2}\frac{R_H^{D-3}}{r^{D-1}}\\
     &-& \frac{(D-3)^2(D-4)(D-2)R_H^{D-5}\lambda'}{4 r^{D-1}}
\end{eqnarray}

Keeping only leading order terms, the deflection angle is
\begin{equation}\notag
\hat{\alpha}=-\int\limits_{0}^{\pi }\int\limits_{\frac{\mathsf{b}}{\sin \phi }%
}^{\infty }\left[ -  \frac{(D-3)(D-2)}{2}\frac{R_H^{D-3}}{r^{D-1}} \right]r dr d\phi
\end{equation}
\begin{equation}
-\int\limits_{0}^{\pi }\int\limits_{\frac{\mathsf{b}}{\sin \phi }%
}^{\infty }\left[ -  \frac{(D-3)^2(D-4)(D-2)R_H^{D-5}\lambda'}{4 r^{D-1}} \right]r dr d\phi. 
\end{equation}
Solving this integral, in the limit $b>>R_H$, and using the relation 
\begin{eqnarray}
    \int_0^\pi \sin^n \phi \,d\phi=\frac{\sqrt{\pi}\, \Gamma\left(\frac{1+n}{n}\right)}{\Gamma\left(\frac{n+2}{2}\right)},
\end{eqnarray}
to the leading order we find
\begin{equation}\notag
\hat{\alpha}\simeq \frac{(D-3)(D-2)}{2}\frac{\sqrt{\pi}\,R_H^{D-3}\, \Gamma\left(\frac{D-2}{D-3}\right) b^{3-D}}{\Gamma\left(\frac{D-1}{2}\right)}
\end{equation}
\begin{equation}
+\frac{(D-3)^2(D-4)(D-2)}{4}\frac{\sqrt{\pi}\,R_H^{D-5}\, \Gamma\left(\frac{D-2}{D-3}\right) b^{3-D}\lambda'}{\Gamma\left(\frac{D-1}{2}\right)} .
\end{equation}

The leading order term agrees with a similar work in Ref. \cite{Belhaj:2020rdb}. Finally, if we perform the rescaling  $\lambda' \to \lambda/(D-4)$ and $D=4$, we obtain
\begin{equation}
\hat{\alpha}\simeq \frac{2 R_H+\lambda/R_H}{b}.
\end{equation}

From the last result it appears that the deflection angle is affected by the string correction, however a closer inspection allows to rewrite in terms of the black hole mass  
\begin{eqnarray}
    \hat{\alpha}\simeq \frac{4 \left(R_H/2+\lambda/4R_H\right)}{b}=  \frac{4M}{b},
\end{eqnarray}
which is identical to the Schwarzschild case. In fact, one can argue that the string correction term is proportional to $\delta\hat{\alpha}_{string} \sim M^2 \lambda/b^4$. Since $b>>M$, practically, this term is very very small. This means that we cannot distinguish the 4D SCBH from the Schwarzschild black hole using the leading term in the weak deflection angle. The small angles lens equation (in the weak deflection approximation) reads
\begin{eqnarray}\label{EinsteinRing}
    \beta=\theta-\frac{D_{LS}}{D_{OS}}\hat{\alpha} .
\end{eqnarray}
In the special situation $\beta=0$, when the source lies on (or passes through) the optical axis an Einstein ring is formed. The weak deflection approximation, $\hat{\alpha} \ll 1$ represents the angular radius of the Einstein ring given by
\begin{eqnarray}\label{EinsteinRing1}
    \theta_{E}\simeq\frac{D_{LS}}{D_{OS}}\hat{\alpha}(b).
\end{eqnarray}
Here we took into account that $D_{OS}=D_{OL}+D_{LS}$, when the angular source position is $\beta=0$. Keeping only the first order of the deflection angle and using the relation $b=D_{OL}\sin{\theta}\simeq D_{OL}\theta$ the bending angle in the small angle approximation is
\begin{eqnarray}
    \theta_{E}\simeq \sqrt{\frac{4M}{D_{OS}}\frac{D_{LS}}{D_{OL}}}.
\end{eqnarray}

It follows that, using only the leading order terms in the weak gravity regime, one cannot distinguish the 4D SCBH geometry from the Schwarzschild by the means of Einstein rings either. 
\begin{figure}
\centering
    \includegraphics[width=3.4 in]{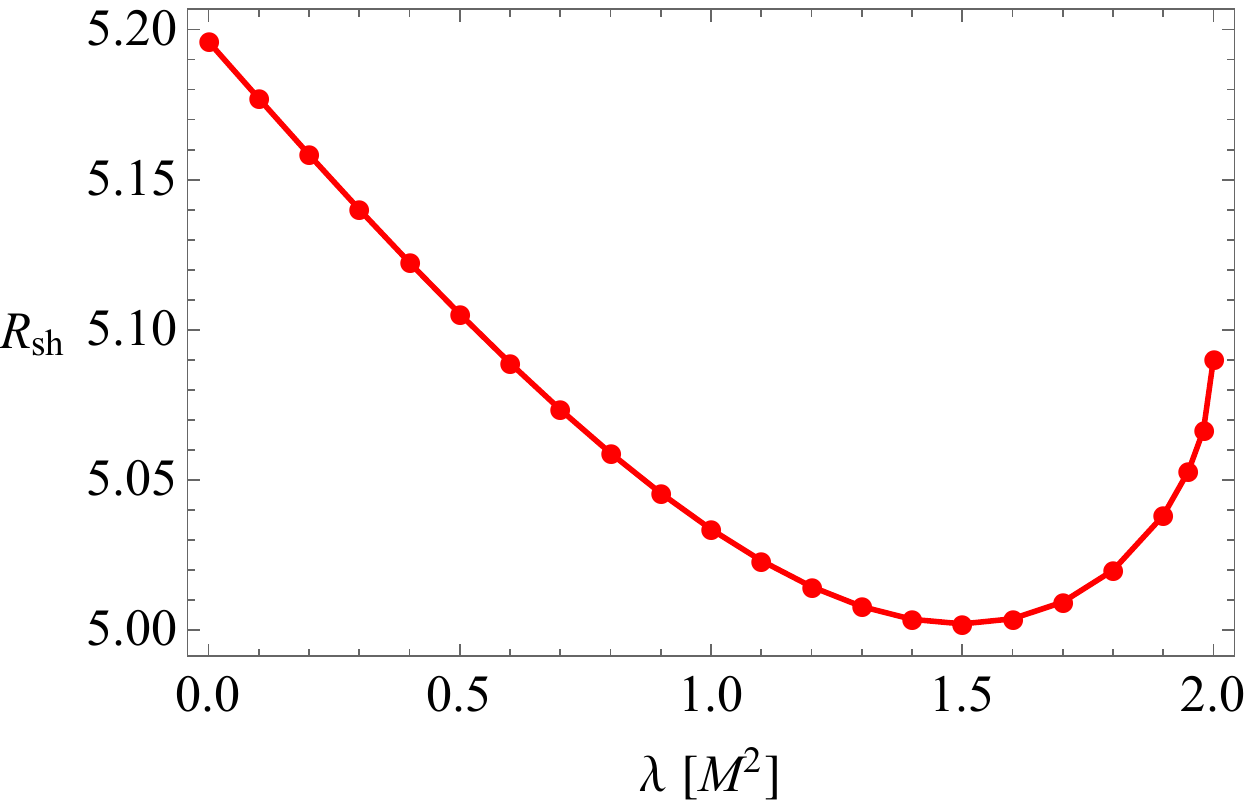}
\caption{Black hole shadow radius as a function of the parameter $\lambda$. We set the black hole mass to $M=1$. \label{fig8}} 
\end{figure}

\begin{figure*}
\centering
    \includegraphics[width=3.1 in]{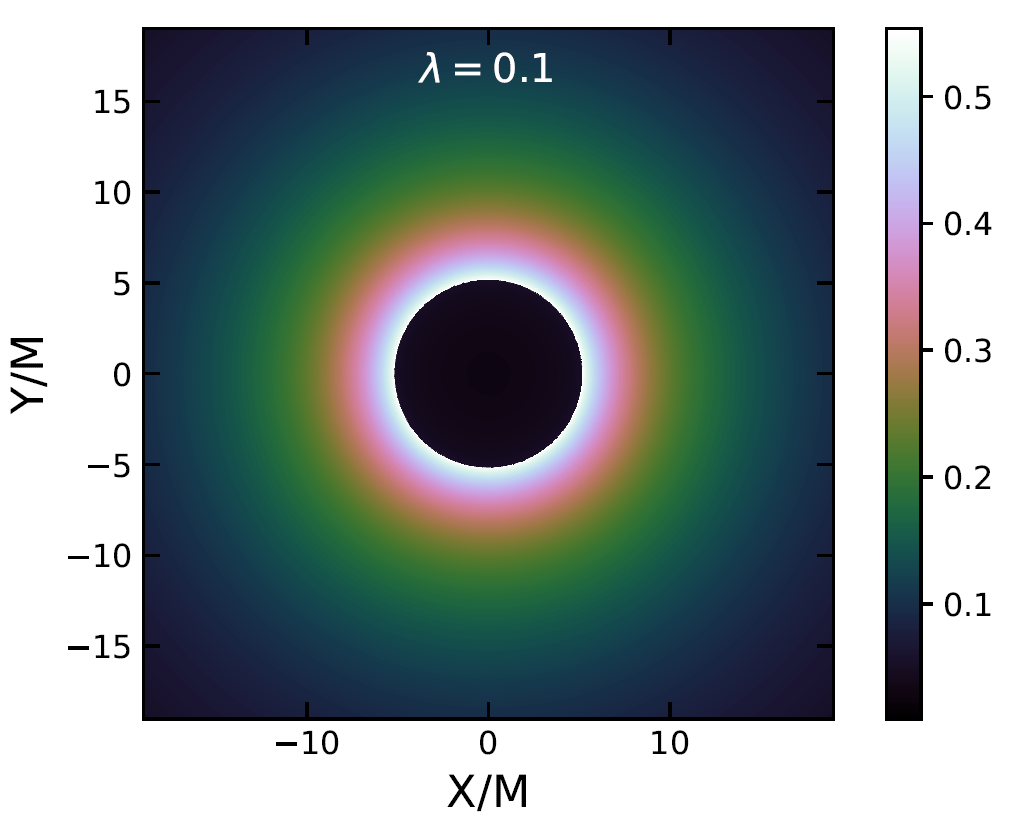}
        \includegraphics[width=3.1 in]{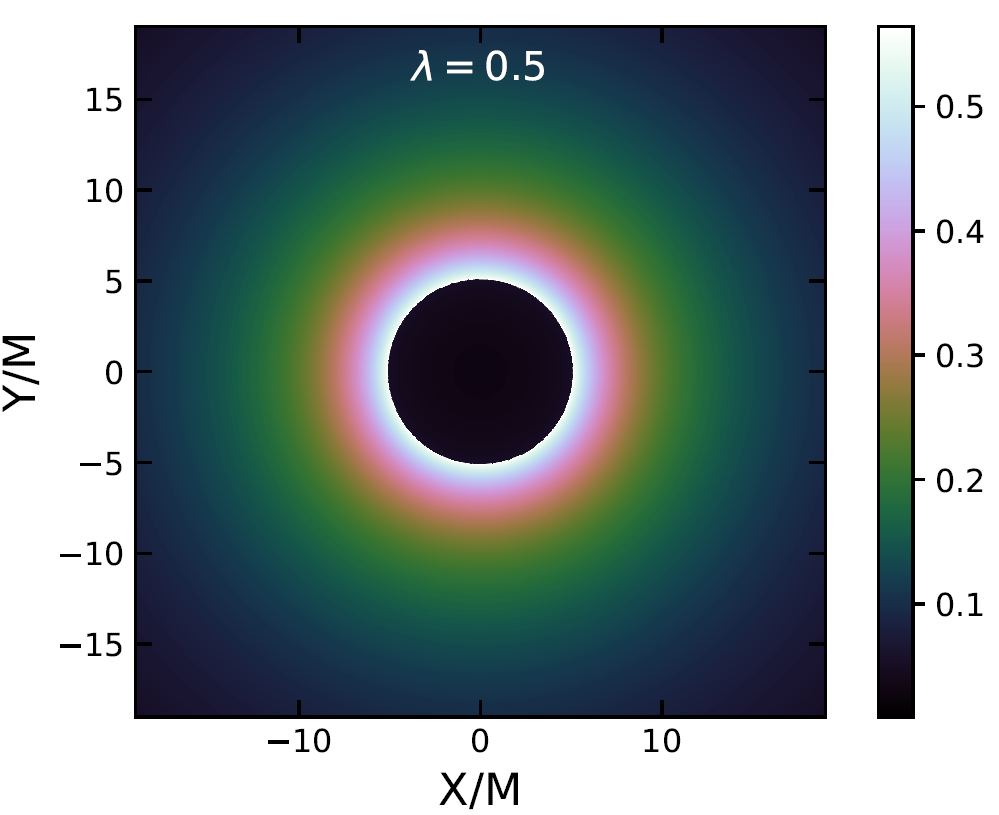}
        \includegraphics[width=3.1 in]{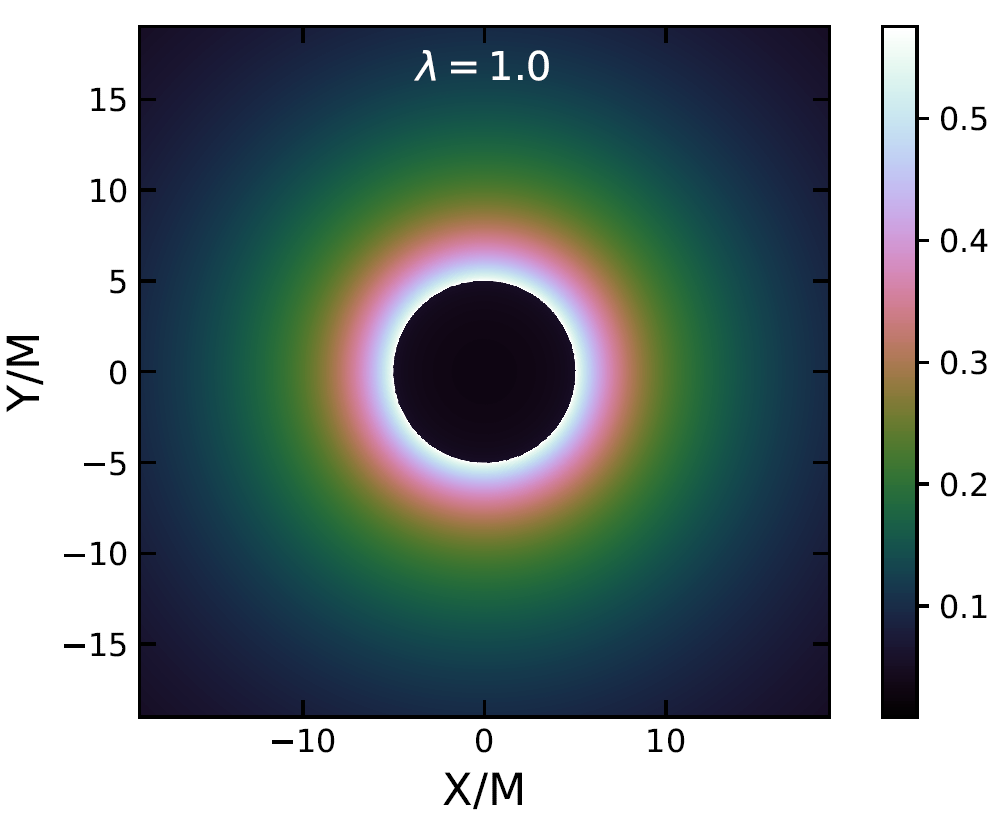}
        \includegraphics[width=3.1 in]{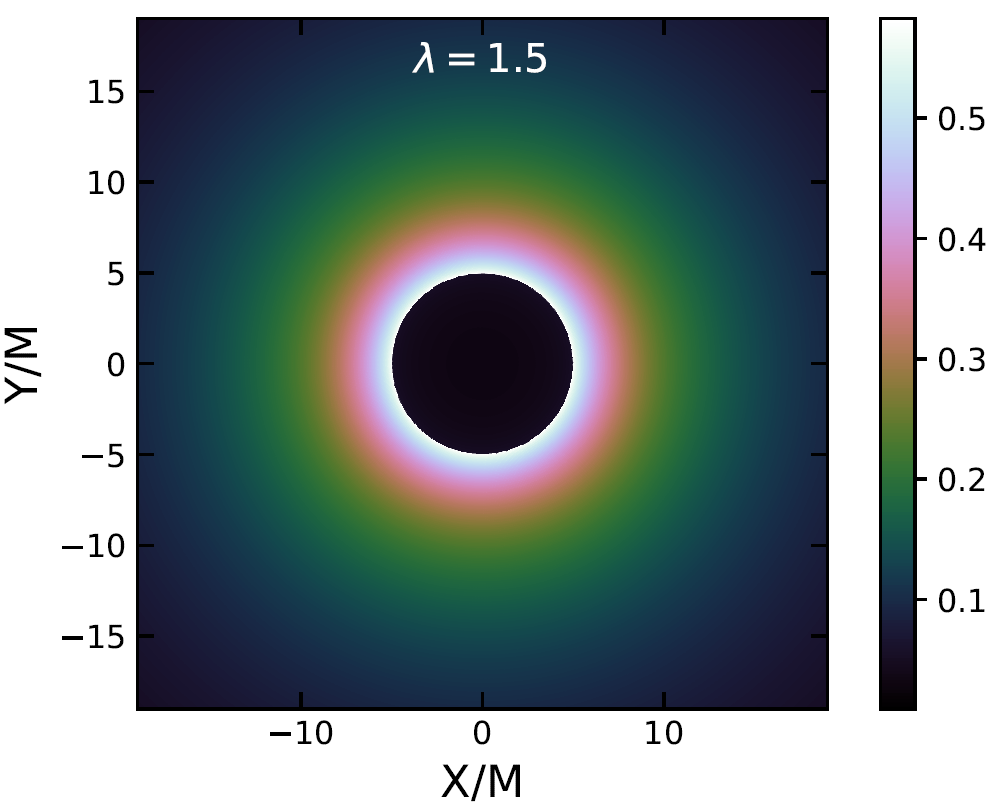}
\caption{Black hole shadow images and the intensities of the 4D SCBH using an infalling gas and radiation model using different values of $\lambda$. We set the black hole mass to $M=1$.   \label{fig9} } 
\end{figure*}


\section{Shadow Images of a 4D SCBH}
From the recent observations using the   
radio images we know that there are supermassive black holes in galactic centers. In particular the black-hole shadows have become a very useful tool to test not only the general relativity in the strong gravity regime but also to probe fundamental physics (see for example \cite{f1,f2,f3,f4,f5,f6,f7}).
Obtaining a realistic image of the black hole is not an easy task. The main reason for this is the complicated surrounding matter near the black hole. In particular, the size, the shape, as well as the composition of the accretion disk are all important. In addition the geometry of magnetic field also influences the shadow \cite{Akiyama1,Akiyama2}. In this work, we will consider a rather simple scenario of having an infalling gas model to obtain the black hole images \cite{Narayan,Saurabh:2020zqg,Zeng:2020dco,Falcke:1999pj,Bambi:2013nla}. For the four-velocity we have \cite{Bambi:2013nla}
\begin{eqnarray}
u^t_{e}  =  \frac{1}{f(r)}, u^r_{e}  =  -\sqrt{\left(1-f(r)\right)}, u^{\theta}_{e}  =u^{\phi}_{e}=  0.
\end{eqnarray}
On the other hand, we have a Hamilton-Jacobi equation
\begin{equation}
\frac{\partial S}{\partial \sigma}+H=0,
\end{equation}
where $S$ is the Jacobi action and $\sigma$ is an affine parameter along
the geodesics.  For the motion of photons in our 4D SCBH we obtain 
\begin{equation}
\frac{1}{2}\left[-\frac{p_{t}^{2}}{f(r)}+f(r)p_{r}^{2}+\frac{p_{\phi}^{2}}{r^{2}}\right] =0,
\label{EqNHa}
\end{equation}
where $p_{t}=-E$ and $p_{\phi}=L$, are two constants of motions; namely the energy and the angular momentum of the photon, respectively. The unstable orbits are characterized by the conditions \cite{Perlick:2015vta}
\begin{equation}
V_{\rm eff}(r) \big \vert_{r=r_{ph}}=0,  \qquad \frac{\partial V_{\rm eff}(r)}{\partial r}%
\Big\vert_{r=r_{ph}}=0.
\end{equation}
From the above equations one can show \cite{Perlick:2015vta}
\begin{equation}
\frac{dr}{d\phi}=\pm r\sqrt{f(r)\left[\frac{r^{2}f(R)}{R^{2}f(r)} -1\right] }. 
\end{equation}
Let us consider a light ray sent from a static observer located at a position $r_{0} $ and transmitted at an angle $\vartheta$ with respect to the radial direction. We therefore have 
\begin{equation}
\cot \vartheta =\frac{\sqrt{g_{rr}}}{g_{\phi\phi}}\frac{dr}{d\phi}\Big\vert%
_{r=r_{0}}.  \label{Eqangle}
\end{equation}
Finally,  the relation for shadow radius of the black hole as observed by a static observer at the position $r_0$  can be written as 
\begin{equation}
R_{sh}=\lim_{r_0 \to \infty}R\sqrt{\frac{f(r_{0})}{f(R)}}\Bigg\vert_{R=r_{ph}}=\frac{r_{ph}}{\sqrt{f(r_{ph})}},
\end{equation}
where $r_{ph}$ represents the unstable photon orbit and  $ r_{0} $ is the position of the observer located at a far distance from the black hole. In Fig. \ref{fig8} we show how the shadow radius changes with the parameter $\lambda$. We observe that the shadow radius decreases initially  in the interval $\lambda \in (0, 1.5)$, it reaches the minimal value (or a reflecting point) at $\lambda/M^2=1.5$, and finally increases in the interval $\lambda \in (1.5, 2)$. The presence of the reflecting point at $\lambda_c$ is explained by the fact the the shadow radius is proportional to the outer horizon. The minimal value of the shadow radius is obtained for $\lambda/M^2=1.5$, having $R_{sh}/M^2 \simeq 5.0021 $ with a photon sphere 
\begin{equation}
    r_{ph}=\frac{1}{2}+\frac{(19+3 \sqrt{33})^{1/3}}{2}+\frac{2}{(19+3 \sqrt{33})^{1/3}}\simeq 2.7589.
\end{equation}

We use a numerical technique known as Backward Raytracing to find the shadow cast by the radiating flow. The total observed flux is given by
\begin{equation}
    F_{obs}(X,Y) =\int_{\gamma} I_{obs}(\nu_{obs},X,Y) d\nu_{obs},
\end{equation}
which can also be written as
\begin{equation}
    F_{obs}(X,Y) =\int_{\gamma}\int g^4 j(\nu_{e})dl_\text{prop}d\nu_{e}.
\end{equation}

We shall assume that our emissivity is monochromatic with the radial profile given by
\begin{equation}
    j(\nu_{e}) \propto \frac{\delta(\nu_{e}-\nu_{\star})}{r^2},
\end{equation}
where $\delta$ is the Dirac delta function. If we use the condition $p_{\mu}p^{\mu}=0$, we can first obtain the relation between the radial and time component of the four momentum, and use the redshift function $g$ given by
\begin{eqnarray}
   g= \frac{k_{\alpha}u^{\alpha}_o}{k_{\beta}u^{\beta}_e}.
\end{eqnarray}
We can then obtain the final expression for the observed flux as follows (for the details see \cite{Bambi:2013nla})
\begin{eqnarray} \label{inten}
    F_{obs}(X,Y) \propto -\int_{\gamma} \frac{\mathrm{g}^3 k_t}{r^2k^r}dr.  
\end{eqnarray}

In Fig. \ref{fig9} we have shown our results for the shadow images and the intensities of the 4D SCBH. In particular we have shown four specific cases; the case with the $\lambda/M^2=\{0.1, 0.5, 1.0,  1.5\}$, where the minimal shadow radius corresponds to the case $\lambda/M^2=1.5$. From the resulting images and their intensities one can see that they look apparently the same in all cases. We can therefore conclude that the 4D SCBH is very difficult to distinguish from the Schwarzschild black hole. Notice also that due to the fact that the deflection angle in leading order terms is the same in units of the black hole mass, this explain why the intensities of the electromagnetic radiation observed far away from the black hole will be almost the same as the Schwarzschild black hole. Such a small contribution is to be expected having in mind our metric (\ref{m23}) according to which the string corrections behave as the inverse of $r^{4}$. 

\section{Conclusions}

In this paper we performed a simple rescaling of the coupling parameter $\lambda$ in the D-dimensional Callan-Myers-Perry black hole to obtain a 4D black hole with string corrections. 
While we lose the direct connection to the original perturbative action, such a rescaling removes divergences at the level of equations of motion, which may give us a glimpse of what string corrections may look like in 4D.

Among the other things, we found that the black hole entropy acquires logarithmic corrections to the usual Bekenstein-Hawking entropy.
We also found a critical value of the coupling parameter for which the Hawking temperature goes to zero. In contrast with the standard scenario where entropy goes to zero as the black hole evaporates completely, our black hole solution reaches a minimal entropy at some critical $\lambda$. For this value of $\lambda$ black holes leave stable remnants. While the black hole geometry still contains the central singularity even with stringy corrections, geodesics of the massive particles and light cannot reach it.  We have also found a close connection between the string corrections and the GUP effects.

We have investigated the possibility of testing such a black hole geometry. We first  constrained its parameters using the S2 star orbit at the galactic center.  We calculated the light deflection angle and the Einstein rings to the leading order terms, as well as the corresponding shadow images using an infalling gas model. While the corrections depend on the string coupling parameter, in general the magnitude of the change is too small to make a distinction from a Schwarzschild black hole. The maximal change occurs at $\lambda/M^2=1.5$, with $R_{sh}=5.0021 [M]$. Compared to the Schwarzschild shadow radius ($3 \sqrt{3} M$), this implies a change $\Delta R_{sh} \simeq 0.194 [M]$. In terms of the angular radius, we can use $\theta_s =R_{sh}M/D$, for Sgr A$^{*}$ black hole with mass $M\simeq 4.1 \times  10^{6}$M\textsubscript{\(\odot\)} and $D =8.3$ kpc, which implies the decrease on the angular size to be $\Delta \theta_s \simeq 1.40$ $\mu$as. Although this is reasonable change, such an effect is out of reach using the current astronomical observations, but it remains an open question if such an effect can be observed in the near future.

\section*{Acknowledgments} K. J. would like to thank P.~G.~S.~Fernandes for his interesting comments on the manuscript. D.S. is partially supported by the US National Science Foundation, under Grant No. PHY-2014021.

\end{document}